\documentclass[acmlarge]{acmart}

\usepackage{xcolor}
\usepackage{graphicx}
\usepackage{float}
\usepackage{longtable}
\usepackage{tabularx}
\usepackage{caption}
\usepackage{subcaption}
\usepackage{gensymb}
\usepackage{diagbox}
\usepackage{algorithm}
\usepackage{amsmath}
\usepackage{dsfont}
\usepackage{algpseudocode}

\AtBeginDocument{%
  }

\setcopyright{none}
\renewcommand\footnotetextcopyrightpermission[1]{}

\begin{document}

\title{Quantum-Assisted Vehicle Routing: Realizing QAOA-based Approach on Gate-Based Quantum Computer}

\author{Talha Azfar}
\email{azfart@rpi.edu}
\orcid{0000-0002-1293-5036}

\author{Osama Muhammad Raisuddin}
\email{raisuo2@rpi.edu}
\orcid{0000-0001-6764-9718}

\author{Ruimin Ke}
\authornote{Corresponding Author}
\email{ker@rpi.edu}
\orcid{0000-0001-9139-6765}

\author{José Holguín-Veras}
\email{jhv@rpi.edu}
\orcid{0000-0001-8118-9383}

\affiliation{%
  \institution{Rensselaer Polytechnic Institute}
  \city{Troy 12180}
  \state{New York}
  \country{USA}
}


\begin{abstract}
The Vehicle Routing Problem (VRP) is a fundamental combinatorial optimization challenge with broad applications in logistics and transportation. In this work, we present a quantum-assisted framework that integrates the Quantum Approximate Optimization Algorithm (QAOA) with a link-based formulation of VRP. Our approach encodes flow conservation and subtour elimination directly into the cost Hamiltonian, preserving graph structure while minimizing resource requirements for practical hardware implementation. We design and implement the full pipeline on a gate-based quantum computer, including problem formulation, encoding, circuit synthesis, and execution on IBM Quantum System One. Experimental results on small VRP instances highlight the effects of penalty scaling, coefficient normalization, and circuit depth on solution feasibility under hardware noise. While scalability remains constrained by circuit complexity and decoherence, the study demonstrates a practical pathway for implementing VRP on quantum hardware and identifies methodological directions for advancing near-term quantum optimization.
\end{abstract}





\maketitle

\renewcommand{\arraystretch}{1.25}
\section{Introduction}

Metropolitan areas generate over 60\% of global GDP but account for nearly 70\% of fossil fuel-related CO$_2$ emissions \cite{NASA_Science_2024}, underscoring the dual challenge of economic growth and environmental externalities. Urban logistics optimization is central to mitigating congestion and inefficiencies while promoting sustainability \cite{he2023brownian}, yet remains complex due to numerous variables and constraints. At its core lies the Vehicle Routing Problem (VRP, a fundamental NP-hard extension of the Traveling Salesman Problem (TSP) \cite{clarke1964scheduling,junger1995traveling}, which seeks efficient vehicle routes to minimize costs such as distance, fuel, or time, and has broad applications from last mile delivery to large scale freight distribution.

VRP instances arise in logistics and transportation planning, from last-mile delivery to large-scale distribution, and even modest-sized instances are NP-hard and computationally intractable to solve optimally with classical algorithms. This motivates exploring alternative computing paradigms for VRP and similar combinatorial problems. In particular, quantum computing has emerged as a promising paradigm for tackling NP-hard problems by exploiting quantum superposition and entanglement to potentially search huge solution spaces more efficiently. Although current quantum devices operate in the noisy intermediate-scale quantum (NISQ) regime with only tens to hundreds of qubits and significant noise, the field is advancing rapidly \cite{preskill2018quantum}. Quantum algorithms for optimization, if carefully designed to work within these hardware limits, could eventually augment classical high-performance computing (HPC) in solving real-world optimization tasks.

\begin{figure}[!htpb]
    \centering
    \includegraphics[width=1\linewidth]{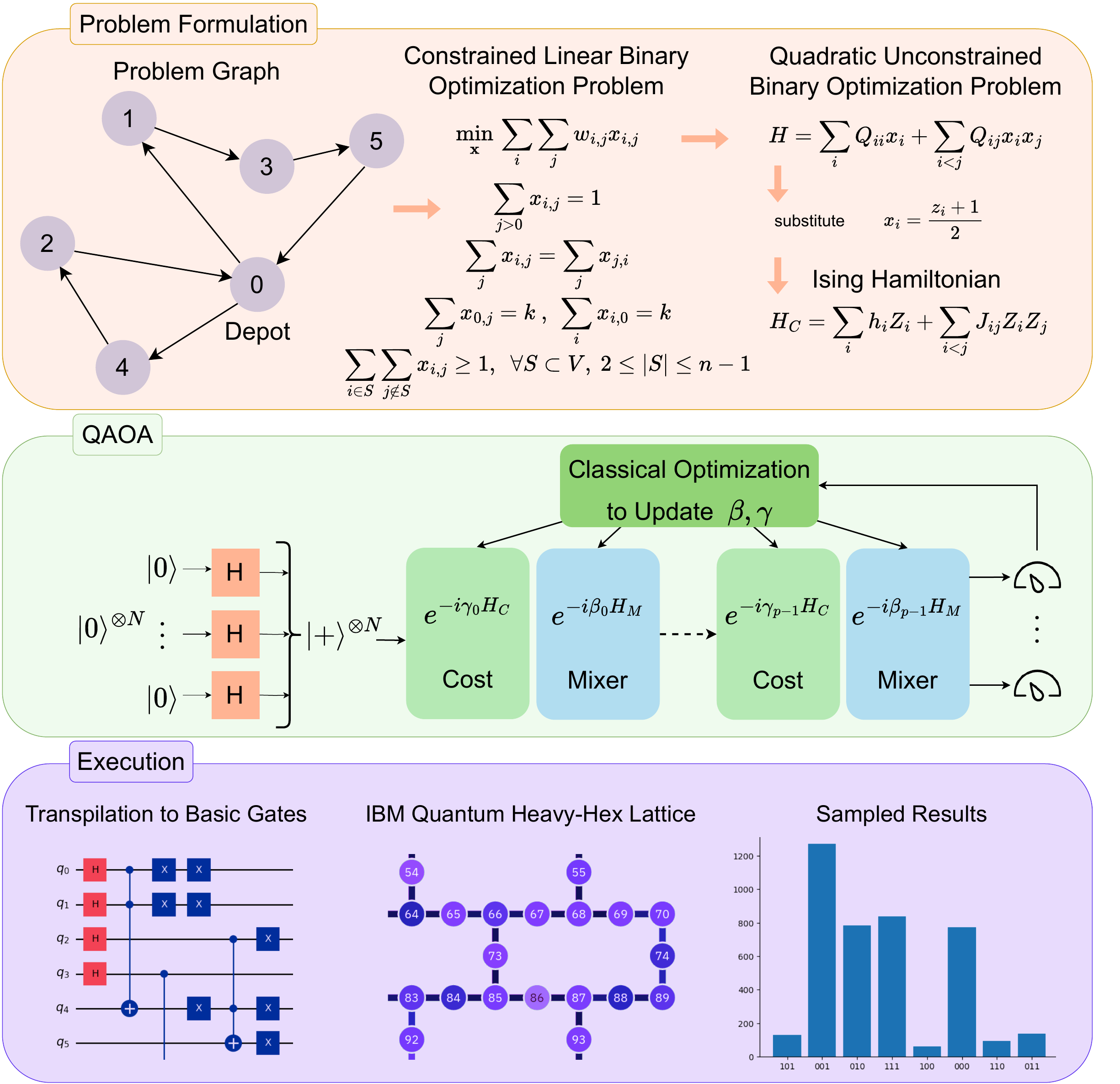}    
    \caption{Solving the VRP using QAOA involves a formulation of the link based constrained minimization form to the unconstrained quadratic binary representation, which is then cast as an Ising Hamiltonian required for QAOA. The high level circuit is transpiled into one and two qubit basic gates for implementation on the IBM-Rensselaer quantum computer. The circuit is run repeatedly to optimize the variational parameters, and a final sampling provides the solution. The problem graph shown here is for illustrative purposes only. Section 4 describes a specific example problem.}
    \label{fig:1}
\end{figure}

One leading approach for combinatorial optimization on gate-based quantum computers is the Quantum Approximate Optimization Algorithm (QAOA) \cite{farhi2014quantum}, inspired by the quantum adiabatic theorem that is fundamental to quantum annealing~\cite{farhi2000quantum}. QAOA is a hybrid quantum-classical algorithm that encodes an optimization problem into a parameterized quantum circuit and uses a classical optimizer to tune those parameters. By alternating problem specific cost operators and mixing operators on $n$ qubits, QAOA can in principle concentrate probability mass on high quality solutions among the $2^n$ possible states. Its suitability for near-term devices has made QAOA a focal point for studying problems in logistics, finance, and network optimization. Applying QAOA to the VRP is especially enticing because VRP's graph-structured decision space might be naturally encoded in a quantum Hamiltonian. A quantum enhanced solution to VRP could not only demonstrate the capabilities of NISQ algorithms on a problem of practical relevance, but also pave the way for quantum assisted HPC workflows where classical and quantum resources are integrated to tackle industry-scale optimization \cite{elsharkawy2025integration}.

However, significant methodological challenges arise in bringing QAOA to VRP. First, one must devise an encoding of VRP into a quantum-friendly cost function (e.g. a Quadratic Unconstrained Binary Optimization, QUBO) that captures all the problem constraints without exploding the resource requirements. The VRP has complex feasibility constraints: each customer must be visited exactly once, vehicles have route length or capacity limits, and routes must form contiguous tours, which are nontrivial to represent in a qubit Hamiltonian. Naively encoding all constraints can require a large number of qubits or high order penalty terms. Prior attempts have taken different formulation strategies, for example, some works introduced time step indexed decision variables to linearize the route order, thereby restricting the search to classically feasible sequences \cite{palackal2023quantum}. This approach can asymptotically reduce the quantum resources needed by eliminating invalid states at the cost of additional ancilla qubits and complex multi-controlled operations. Another approach is to use unbalanced penalty weights for soft constraints, effectively biasing the cost function to discourage violations without strictly forbidding them; but this requires careful tuning of penalty coefficients via classical optimization and can introduce higher order interactions \cite{montanez2024unbalanced}. Designing an encoding that is compact yet enforces VRP constraints remains a core challenge.

Ensuring solution feasibility is a key challenge when applying QAOA to constrained problems. Since the quantum state explores superpositions of many bitstrings, an unsuitable Hamiltonian or mixer can increase the likelihood of measuring infeasible routes (e.g., routes omitting customers or forming subtours). To address this, researchers have explored customized mixers that restrict the evolution to feasible subspaces, as well as warm-start strategies that initialize QAOA in valid solutions. However, empirical studies on VRP have shown only limited improvements from these techniques \cite{fitzek2024applying}. For example, even in hybrid approaches where classical clustering was used to reduce the problem before applying QAOA, performance remained behind comparable VQE methods, with QAOA struggling to consistently recover feasible solutions in small instances. These findings highlight the ongoing difficulty of steering QAOA effectively toward the feasible region of combinatorial problems.

Furthermore, quantum circuit depth and hardware limitations impose severe restrictions on solving VRP with QAOA. Each layer of QAOA applies numerous two qubit gates to implement problem interactions. As the problem size or number of QAOA layers increases, the circuit quickly becomes deeper and more error-prone. Current superconducting quantum processors have limited qubit connectivity and coherence times, meaning complex QAOA circuits suffer from accumulated gate errors and decoherence before completion \cite{blekos2024review}. For a VRP instance of even modest size, achieving good solution quality might require many layers of QAOA, a regime simply not attainable on today's hardware. Moreover, qubit count and topology are limiting factors: encoding a VRP with $N$ locations typically involves $O(N^2)$ binary decision variables (for example, representing each possible directed edge or each assignment of a location to a route position). This can easily exceed the number of physical qubits available, and even when it does not, mapping the logical interactions onto a real device  (such as IBM's heavy-hex lattice) can require additional swap gates that further increase the effective circuit depth.

This work presents a hardware grounded study of QAOA for the Vehicle Routing Problem (VRP), with novelty arising from the co-design of problem formulation, circuit construction, and experimental evaluation on real quantum hardware rather than simulation-only analysis. The main contributions are:
\begin{itemize}
 
    \item We develop a directional edge-based formulation that preserves the graph structure of VRP and embeds flow conservation and subtour elimination directly into the cost Hamiltonian. Unlike time-expanded or route-order encodings, this approach avoids explicit route sequencing variables and reduces qubit and gate overhead, making it more suitable for near-term devices.
 
    \item  We implement the full workflow, from VRP encoding and QUBO construction to Ising mapping, circuit synthesis, transpilation, and execution on IBM Quantum System One. This provides empirical insight into QAOA behavior under realistic noise, connectivity, and depth constraints, which are not captured by idealized simulations.

    \item We quantitatively evaluate how VRP formulations impact qubit count and two-qubit gate depth after hardware-aware transpilation, showing that circuit depth is a primary bottleneck for constrained routing problems on current superconducting architectures.

    \item We demonstrate that both penalty parameter selection and coefficient normalization critically influence solution feasibility. In particular, penalties on the order of twice the sum of edge weights and normalization of Hamiltonian coefficients significantly improve feasible-solution sampling on larger instances.
\end{itemize}

Overall, the novelty of this work lies in its practical, hardware executed evaluation of QAOA for VRP, offering concrete design insights that clarify current limitations and inform future quantum optimization studies in the NISQ regime.

\section{Literature Review}

A universal quantum computer refers to a system that can implement any quantum algorithm and perform arbitrary quantum computations similar to how a Turing machine can execute any classical algorithm, an idea put forth by Richard Feynman~\cite{feynman1982simulating}. A gate based quantum computer can simulate any unitary evolution efficiently using a set of universal quantum gates capable of approximating the unitary operation to arbitrary precision. The theoretical foundations for quantum computation were laid down in \cite{deutsch1985quantum} showing that certain tasks involving randomness can be performed by a quantum computer faster than classical probabilistic algorithms on a classical computer. Some examples of known quantum advantage are factorization using Shor's algorithm~\cite{shor1999polynomial}, unstructured search using Grover's algorithm \cite{grover1996fast}, and simulation of the electronic structures of molecules or other quantum systems \cite{tacchino2020quantum}. 

Special-purpose quantum computers such as quantum annealers are built to exploit quantum mechanics for specific problem domains. Similar to thermal simulated annealing~\cite{kirkpatrick1983optimization}, quantum annealing is a process that finds the lowest-energy configuration of a system by gradually transforming from a simple quantum state to one representing the optimal solution \cite{finnila1994quantum}. It was shown to outperform classical methods for the traveling salesperson problem \cite{martovnak2004quantum}, and has since found commercial use through companies such as D-Wave that are now offering devices with over 5000 qubits which see use in logistics, finance, and machine learning \cite{boothby2020next}. However, special-purpose computers exhibit a speedup that is problem-dependent, and can only perform a narrow set of tasks and lack the flexibility to run any general quantum algorithm \cite{yarkoni2022quantum}. 
The capacitated vehicle routing problem has been approached by a hybrid algorithm utilizing quantum annealing, with the results showing no clear benefit over classical solvers \cite{feld2019hybrid}. This method used a cluster-then-route approach which loses global connections, and subsequent routing cannot be guaranteed to give the true optimum.

A Variational Quantum Algorithm (VQA) is a hybrid quantum-classical approach for universal quantum computing that leverages quantum circuits parameterized by classical variables to solve optimization problems. For example, the Variational Quantum Eigensolver (VQE) seeks to find the minimum eigenvalue or the ground state energy of a Hamiltonian, and has seen application in materials science and quantum chemistry \cite{mcclean2016theory}.
Another VQA, the Quantum Approximate Optimization Algorithm (QAOA) was introduced for approximate combinatorial optimization that aims to find the bit string which maximizes a cost function~\cite{farhi2014quantum}. It guarantees the optimization in the limit of infinite layers in a manner similar to the adiabatic evolution used in quantum annealing~\cite{farhi2000quantum}.

Beyond its use for solving optimization problems, QAOA may be a route to study establishing quantum supremacy since classically emulating QAOA in an efficient way would collapse the polynomial hierarchy of computational complexity. However this is believed to be unlikely \cite{farhi2016quantum}. 
At low depth proportional to $\log n$, the performance of QAOA is known to be low as the algorithm does not see the whole graph in Max-Cut type problems \cite{farhi2020quantum}. 
There are diminishing returns towards accuracy versus QAOA circuit depth \cite{blekos2024review}. In fact, a study into the classical optimizers and circuit depth on noisy quantum devices concluded that beyond 6 to 8 layers the probability of success decreases due to noise accumulation and decoherence \cite{pellow2024effect}. The upper bound on QAOA is yet to be determined in practical scenarios but it has been observed that it outperformed the best classical solvers for large boolean satisfiability problems \cite{boulebnane2024solving}. Importantly, using the QAOA outputs as guesses to the classical optimizer did not yield a significantly better overall performance. 

While interest in quantum computing for intelligent transportation systems has been expressed only recently~\cite{burkacky2020will}\cite{zhuang2024quantum}, the combinatorial optimization problems of TSP and VRP, have received significant attention from a quantum computing perspective \cite{martovnak2004quantum}\cite{leonidas2023qubit}. 
Classical methods exist to solve the problem exactly, for example branch and bound \cite{theurich2021branch}, which scale exponentially with problem size. On the other hand metaheuristic algorithms can provide approximate solutions, like Tabu search \cite{sicilia2016optimization} and genetic algorithms~\cite{baker2003genetic}, but they are not guaranteed to find the exact optimal solution in large problems. 
QAOA was applied to small scale TSP instances, giving high quality solutions at limited depth with a mixer Hamiltonian corresponding to the problem constraints \cite{ruan2020quantum}. The performance of QAOA simulation on a VRP formulation without subtour elimination constraints was found to depend strongly on the classical optimization method and the depth of the ansatz since training time can increase exponentially, but this may be prevented by a good initialization~\cite{azad2022solving}. However, the authors acknowledge the appearance of subtours in the QAOA results and recommend explicit penalization. Another study discovered that QAOA biases solutions toward lower energies but fails to distinguish between feasible solutions of different costs, limiting its effectiveness for VRP \cite{fitzek2024applying}. The authors conjecture that this may be due to a small energy gap in the Hamiltonian representation, which could be mitigated by reformulating the problem to increase energy separation. A detailed comparison of different VRP formulations and quantum algorithms demonstrated that QAOA with constrained optimization by linear approximation (COBYLA) delivered the highest probability of sampling the optimal configuration given enough post optimization sampling shots \cite{harwood2021formulating}. 

Using a hybrid approach of clustering on a classical computer and then solving the subproblem of capacitated VRP using QAOA and VQE was tested in \cite{palackal2023quantum}. Despite the time expanded formulation, VQE achieved feasible solutions on small instances, QAOA consistently underperformed, and extensions such as constraint-preserving mixers and warm-start QAOA did not improve results. Similar approaches that incorporate time steps into the decision variables and restrict the state space to feasible bitstrings achieve asymptotically lower resource requirements, but in practice incur substantial ancilla qubit overhead for small instances \cite{xie2024feasibility}. Other approaches to simplify the quantum circuit involve the use of unbalanced penalization for inequality constraints \cite{montanez2024unbalanced}, however, these require classical optimization to determine the penalty coefficients. 

In this work, we formulate the VRP using directional links as decision variables with subtour elimination constraints, and investigate the implementation of QAOA, focusing on the encoding strategy, circuit depth, and scalability on a real quantum computer.

\section{Conceptual Overview}
\subsection{Quantum Computing}
A qubit (quantum bit) is the fundamental unit of quantum information in a quantum computer. Unlike classical bits, which can be in a state of 0 or 1, a qubit can exist in a superposition of both states. While the state of a qubit cannot be observed directly, repeated measurements can be performed to estimate the probability distribution.  A qubit can be represented as:
\begin{equation}|\psi\rangle=\alpha|0\rangle+\beta|1\rangle \label{eq:qubit}\end{equation}
where $\alpha$ and $\beta$ are probability amplitudes, complex numbers satisfying the completeness relation $|\alpha|^2+|\beta|^2=1$. The notation $|\cdot\rangle$ is called a `ket' which can be represented as a column vector, so $|0\rangle = \begin{bmatrix}1\\0\end{bmatrix}$, and $|1\rangle =  \begin{bmatrix}0\\1\end{bmatrix}$. The `bra' notation, $\langle \cdot|$, is the adjoint of ket and can be represented as a row vector. 

The state of multiple qubits lies in an exponential state space. If there are $n$ qubits, each of the $2^n$ basis states has a complex coefficient, $\alpha_i$, that dictates its probability, $|\alpha_i|^2$, of measuring that basis state. It is for this reason that quantum computing cannot be efficiently simulated on classical computers \cite{bremner2011classical}. For example, 50 qubits will require $2^{50}$ complex numbers to describe the state, and any circuit operations will require multiplication with a matrix of size $2^{50} \times 2^{50}$. If each coefficient was represented with just one byte, the state vector would take over 1000 Terabytes to store in a classical computer.

The IBM Eagle type quantum computer uses transmon qubits which are a type of superconducting charge qubit, similar to the Cooper-pair box. At a fundamental level the transmon qubit works on the principle of the Josephson effect~\cite{josephson1964coupled}, which is the appearance of a constant current through an insulating barrier between two superconductors in the absence of an external electromagnetic field due to quantum tunneling. This effect enables quantum coherence, and combined with precise microwave control, makes the transmon a stable and controllable qubit for quantum computing \cite{koch2007charge}. Microwave radiation can induce quantized voltages across the junction, allowing control, coupling, and measurement using microwave resonators. The qubit states $|0\rangle$ and $|1\rangle$ are typically assigned to the two lowest energy levels of the transmon. There also exist physical coupling channels on the quantum device that allow direct capacitive coupling between qubits to implement two-qubit entangling gates. 

The evolution over time of a quantum state $|\psi(t)\rangle$ of $n$ qubits is described by a differential equation known as the Schr\"{o}dinger equation,
$$i\hslash \frac{d}{dt}|\psi(t)\rangle = \mathbf{H}|\psi(t)\rangle$$

where $i=\sqrt{-1}$, $\hslash$ is the reduced Planck's constant, and $\mathbf{H}^{2^n \times 2^n}$ is a Hermitian matrix known as the Hamiltonian. The solution of the equation yields an exponential operator $U(t) = e^{i\mathbf{H}t/\hslash}$, for any $t>0$, 
$$|\psi(t)\rangle  = U(t)|\psi(0)\rangle  $$

The $U$ matrix corresponds to rotations of the qubits implemented as quantum gates on the hardware. An operation on $n$ qubits can be represented by a unitary matrix, $U\in \, \mathds{C}^{2^{n} \times 2^{n}}$. It satisfies the following properties: 
$$U^{-1} = U^\dagger, \ \   UU^\dagger = U^\dagger U = I$$

where $^\dagger$ refers to the Hermitian conjugate (complex conjugate transpose) and $I$ is the identity matrix. 

These operations can transform a known quantum state to any desired state in a quantum computer. One basic gate is the Hadamard gate, which, when applied to a qubit in the zero state, converts it into an equal superposition of basis states $|0\rangle$ and $|1\rangle$.
$$H = \frac{1}{\sqrt 2}\begin{bmatrix} 1 & 1 \\ 1 & -1\end{bmatrix}, \ \ H|0\rangle = \frac{1}{\sqrt 2} \begin{bmatrix} 1 & 1 \\ 1 & -1\end{bmatrix} \begin{bmatrix} 1 \\ 0 \end{bmatrix} = \frac{1}{\sqrt 2}\begin{bmatrix}1 \\1\end{bmatrix}$$

Applying it to $|1\rangle$ creates an equal superposition as well, but the signs of the probability amplitudes differ. This characteristic of probability amplitudes allows quantum states to interfere, which is not possible in classical stochastic systems which are described using probabilities. 
$$H|0\rangle=\dfrac{|0\rangle+|1\rangle}{\sqrt{2}}, \quad H|1\rangle=\dfrac{|0\rangle-|1\rangle}{\sqrt{2}}$$

\begin{figure}[ht!]
    \centering
    \begin{subfigure}[b]{0.35\linewidth}
        \includegraphics[width=\textwidth]{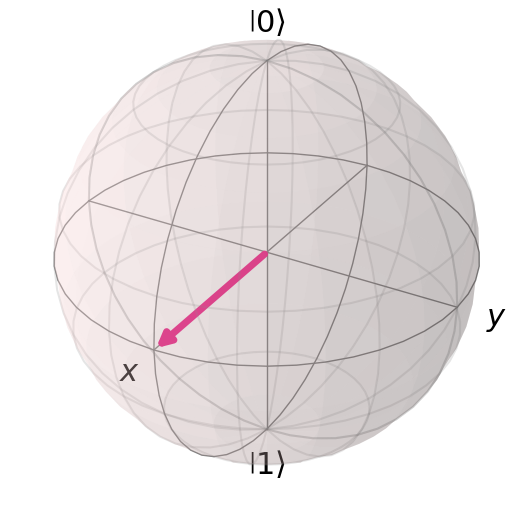}
        \caption{$|+\rangle = H|0\rangle$}
        \label{fig:subfig1}
    \end{subfigure}
~    \begin{subfigure}[b]{0.35\linewidth}
        \includegraphics[width=\textwidth]{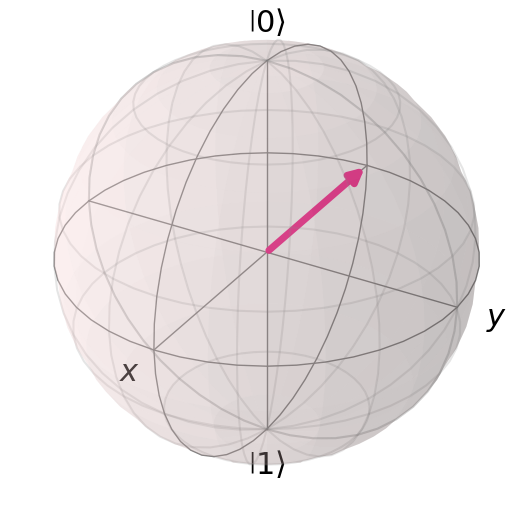}
        \caption{$|-\rangle = H|1\rangle$}
        \label{fig:subfig2}
    \end{subfigure}
    \caption{Two states showing even superposition on the Bloch sphere. The Z direction is upwards. }
    \label{fig:blochs}
\end{figure}

If measured on the computational basis $|0\rangle,|1\rangle$ on a quantum computer, the samples of 0s and 1s from these states would be equal, subject to noise. The state of a qubit can also be depicted on the Bloch spheres in Figure \ref{fig:blochs}, which show the two Hadamard states pointing to opposite points on the equator, each representing an equal  measurement probability between the two poles of the computational basis on the Z axis. The Bloch sphere representation can be derived from Equation \ref{eq:qubit}, 
$$\alpha = \cos{\frac{\theta}{2}}, \ \beta = e^{i\phi}  \sin{\frac{\theta}{2}}$$

where the polar angle, $\theta$, indicates the superposition, and the azimuthal angle, $\phi$, is the relative phase.

Pauli Gates ($X,Y,Z$) Constitute the fundamental quantum operations that rotate a qubit around their respective axes on the Bloch sphere, analogous to classical bit-flip operations.
$X = \begin{bmatrix}0 & 1 \\ 1 & 0\end{bmatrix}$ (Pauli-X) functions similar to a NOT gate: $X|0\rangle=|1\rangle$ , $X|1\rangle=|0\rangle$. 
$Y = \begin{bmatrix}0 & -i \\ i & 0\end{bmatrix}$ (Pauli-Y) is a $\pi$ radian rotation about the Y axis: $Y|0\rangle=i|1\rangle$ , $Y|1\rangle=-i|0\rangle$. 
$Z= \begin{bmatrix}1 & 0 \\ 0 & -1\end{bmatrix}$ (Pauli-Z) flips the phase: $Z|0\rangle=|0\rangle, Z|1\rangle=-|1\rangle$.

\begin{figure}
    \centering
    \includegraphics[width=0.5\linewidth]{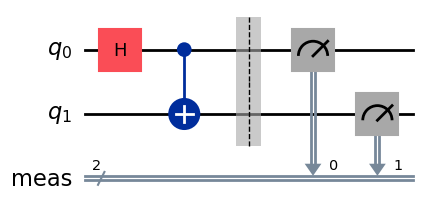}
    \caption{Quantum circuit for creating the Bell state. Qubits start in the $|00\rangle$ state. The element H is a Hadamard gate applied to qubit-0, while the blue symbol with $+$ sign indicates a CNOT gate, with $q_0$ controlling the NOT operation on $q_1$. The two gray elements on the right depict measurement.}
    \label{fig:bell}
\end{figure}

Gates of more than one qubit can result in entanglement. The controlled-not (CNOT) gate is a two qubit gate given by
$$CNOT = \begin{bmatrix}1&0&0&0\\ 0&1&0&0\\ 0&0&0&1\\ 0&0&1&0 \end{bmatrix}$$

which flips the second qubit if the first one is 1. If the first qubit is in the Hadamard state as shown in Figure \ref{fig:bell}, the controlled qubit does not simply inherit the superposition, in fact the two qubits become entangled into the superposition known as the Bell state:
$$\left|\Phi^{+}\right\rangle=\frac{|00\rangle+|11\rangle}{\sqrt{2}}$$

If we measure one qubit as $|0\rangle$, the other qubit is guaranteed to also be in the state $|0\rangle$. These two qubits are in a superposition of two states out of a possible four, meaning that the probability of measuring $|01\rangle$ or $|10\rangle$ is zero.

It is important to understand the intermediate steps. When the Hadamard gate is applied to $q_0$, the matrix that is multiplied to the 4-dimensional state vector is a Kronecker product:
\begin{equation*}
\begin{split}
|\phi\rangle = H \otimes I |00\rangle = \frac{1}{\sqrt{2}} \begin{bmatrix} 1&0&1&0\\ 0&1&0&1\\ 1&0&-1&0\\ 0&1&0&-1\end{bmatrix}\begin{bmatrix}1\\0\\0\\0\end{bmatrix} \\
|\phi\rangle =\frac{1}{\sqrt{2}} \begin{bmatrix}1\\0\\1\\0\end{bmatrix} = \frac{1}{\sqrt{2}} \begin{bmatrix}1\\0\\0\\0\end{bmatrix} + 0 \begin{bmatrix}0\\1\\0\\0\end{bmatrix} + \frac{1}{\sqrt{2}} \begin{bmatrix}0\\0\\1\\0\end{bmatrix} + 0 \begin{bmatrix}0\\0\\0\\1\end{bmatrix} \\
\Rightarrow \ |\phi\rangle =\frac{1}{\sqrt{2}} \left( |00\rangle + |10\rangle  \right)  \\
\end{split}
\end{equation*}

$$|\Phi^+\rangle = CNOT|\phi\rangle = \frac{1}{\sqrt{2}} \begin{bmatrix}1&0&0&0\\ 0&1&0&0\\ 0&0&0&1\\ 0&0&1&0 \end{bmatrix} \begin{bmatrix}1\\0\\1\\0\end{bmatrix}  =  \frac{1}{\sqrt{2}}  \begin{bmatrix}1\\0\\0\\1\end{bmatrix}$$

Quantum computing achieves speedups by employing these principles of superposition, entanglement, and quantum parallelism making it faster for specific problems where classical methods scale poorly.

\subsection{Quantum Approximate Optimization Algorithm}
Quantum Approximate Optimization Algorithm (QAOA) is a hybrid quantum-classical algorithm designed for combinatorial optimization problems by leveraging quantum computing properties on a near-term quantum device in the NISQ era \cite{farhi2016quantum}. QAOA is particularly suited for finding approximate solutions to NP-hard problems such as Max-Cut, traveling salesperson, and graph partitioning. QAOA is inspired by Quantum Annealing, which relies on the Adiabatic Theorem, which states that a system in its ground state (lowest energy) will remain there if the Hamiltonian changes slowly enough. QAOA approximates this gradual evolution using a parameterized circuit derived by Farhi et al. \cite{farhi2014quantum}. It works by encoding the problem into a cost Hamiltonian, then evolving a quantum state using parameterized quantum gates corresponding to the cost Hamiltonian $H_C$ and a mixing Hamiltonian $H_M$ which prepares the state 
\begin{equation}|\psi(\gamma, \beta)\rangle=\prod_{j=1}^p e^{-i \beta_j H_M} e^{-i \gamma_j H_C}|+\rangle \label{qaoa}\end{equation}

where $|+\rangle$ is the equal superposition state (a common choice for the initial state which must be the ground state of the mixing Hamiltonian), $p$ is the number of repetitions, and $\psi$ is a parameterized state. The parameters $(\gamma, \beta)$ are optimized by a classical method based on the expected value of the cost. Once the parameters converge, the circuit can be sampled numerous times to find the solution as the most likely state. Since QAOA provides approximate solutions it is more suitable at larger scales when it becomes infeasible to solve on classical computers. While exponential speedup cannot be guaranteed, it has been shown to outperform classical heuristics in certain cases \cite{zhou2020quantum}. 

The optimization problem has to be expressed in the form of a cost Hamiltonian with no constraints. Many combinatorial problems can be naturally expressed as a quadratic unconstrained binary optimization (QUBO), which encodes optimization problems using binary variables. A key advantage of QUBO is that it can be directly mapped to the Ising Hamiltonian, a fundamental model in quantum mechanics used to describe interactions in spin systems. It allows for efficient Trotterization \cite{kluber2023trotterization}, enabling simulation of time evolution using simple quantum gates. Additionally, it makes expectation values easy to estimate via repeated measurements in the computational basis. The Ising model is particularly relevant in quantum optimization, as it forms the basis for quantum annealing making it compatible with QAOA as a cost Hamiltonian. A QUBO problem is generally expressed as:
$$\min \, \mathbf{x}^T\mathbf{Qx}$$

where $\mathbf{x}$ is an $n$ bit long binary vector and $\mathbf{Q}$ is a matrix of real valued coefficients. The cost function can be split into linear and quadratic terms since $x_i^2=x_i$,
\begin{equation}C(\mathbf{x})=\sum_i Q_{i i} x_i+\sum_{i<j} Q_{i j} x_i x_j \label{eq:Hc}\end{equation} 

To map the problem to the Ising Hamiltonian, the problem must be defined in terms of spin variables which take the values as $\pm 1$. This corresponds to the direction of the Z axis on the Bloch sphere, which is the computational basis. This leads to the conversion by substituting

$$x_i = \frac{1}{2}(z_i+1) ,$$
\begin{equation}C(\mathbf{z})=\sum_i \frac{Q_{i i}}{2}\left(z_i+1\right)+\sum_{i<j} \frac{Q_{i j}}{4}\left(z_i z_j+z_i+z_j+1\right)\label{eq:Ising}\end{equation}

Spin interactions correspond to Pauli-Z operator being applied on the qubits. The expression can be rearranged to give
\begin{equation}
    C(\mathbf{x}) = C(\mathbf{z}) = \langle \mathbf{x}|H_C|\mathbf{x} \rangle
\end{equation}
where $H_C$ is the cost Hamiltonian operator
\begin{equation}
    H_C=\sum_i h_i Z_i+\sum_{i<j} J_{i j} Z_i Z_j + \text{constant terms} \label{eq:Hising}
\end{equation}

for some real coefficients $h_i$, and $J_{ij}$,  where $Z_i$ is the single qubit Pauli-Z gate acting on the $i$th qubit, while $Z_iZ_j$ refers to a two qubit Pauli-Z gate defined as the tensor product
$$Z \otimes Z=\begin{bmatrix}1 & 0 & 0 & 0 \\ 0 & -1 & 0 & 0 \\ 0 & 0 & -1 & 0 \\ 0 & 0 & 0 & 1\end{bmatrix}$$

In an $n$ qubit system the operation $Z_i$ can be expressed as $\dotsi   I_{i-1} \otimes Z_i \otimes I_{i+1} \dotsi$, and $Z_iZ_j = \dotsi  I_{i-1} \otimes Z_i \otimes  I_{i+1} \dotsi \otimes I_{j-1} \otimes Z_j \otimes I_{j+1} \dotsi$.

The QAOA can now be executed. First the Hadamard gate is used to create a uniform superposition across all states
$$|\psi(0)\rangle = H|0\rangle = \frac{1}{\sqrt{2^n}} \sum_{x=0}^{2^n-1}|x\rangle = |+\rangle $$

The state is evolved under the cost Hamiltonian for a duration $\gamma_0$, which is a tunable parameter
\begin{equation}U_C(\gamma_0)=e^{-i \gamma_0 H_C} \label{U_C}\end{equation}

This is followed by the application of a mixing Hamiltonian for a duration of $\beta_0$
\begin{equation}H_M = \sum_i X_i \,, \ \ U_M(\beta_0) = e^{-i\beta_0 H_M} \label{U_M}\end{equation}

The choice of $Z$ and $ZZ$ for the cost and $X$ for the mixer is due to the requirement that both Hamiltonians do not commute. The cost and mixing Hamiltonians are repeatedly applied for some $p \ll 2^n $ times for different values of $\gamma_i$ and $\beta_i$. The optimum values of these parameters has to be tuned using classical optimization methods, such as gradient descent. 

After the optimization, the states sampled from $|\psi(\gamma,\beta) \rangle$ are likely to approximate the optimal solution of the optimization problem. The approximation ratio $r$ is the expected value of the cost function divided by the optimum. A single layer QAOA for Max-Cut can achieve $r \approx0.6924 $, and theoretically $r$ approaches 1 as depth increases. However, due to noise and other practical limitations deep QAOA circuits are difficult to implement, but still prove to be better than the best classical SAT solvers \cite{boulebnane2024solving}. It has been conjectured that QAOA may be able to provide exponential speedup in certain cases \cite{montanez2024towards}.

\subsection{Vehicle Routing Problem}
An instance of the capacitated VRP is characterized by a set of customers with specific demand, a fleet of vehicles with limited capacity, a central depot where vehicles start and end their routes, and a matrix representing travel costs between locations. In this study, we assume that customer demand is satisfied upon the vehicle's arrival at the designated node. Therefore, the problem becomes to minimize the sum of the distance of all links traversed. If we denote the binary variable indicating a directional link between nodes $i$ and $j$, in the set of nodes $\mathcal{N}$, being active or not as $x_{i,j}$, with the distance or cost in the link as $w_{i,j}$, the objective for the link based formulation is expressed as:
\begin{equation}\min_{\mathbf{x}} \, \sum_i\sum_j w_{i,j}x_{i,j} \ \quad i,j\in \mathcal{N}, \label{eq:vrp} \end{equation}

subject to the following constraints: that each customer is visited exactly once,
$$\sum_{j>0}x_{i,j} = 1 \,  \quad  \forall i>0 \  ,$$

if a vehicle enters a node, it must leave,
$$\sum_{j} x_{i,j} = \sum_{j} x_{j,i} \  \quad \forall i \ ,$$

and there are exactly $k$ vehicles leaving and entering the depot,
$$\sum_{j}x_{0,j} = k \ , \ \sum_{i}x_{i,0} = k \ .$$

Additionally, in order to prevent disjoint loops that do not visit the depot, subtour elimination constraints must be imposed,
$$\sum_{i \in S} \sum_{j \notin S}x_{i,j} \geq 1, \quad \forall S \subset \mathcal{N}, \ 2 \leq |S| \leq n-1$$

where $S$ is any subset of the node set $\mathcal{N}$, consisting of $n$ total nodes. Meaning that any selected subset of the nodes must have at least one active edge to some node in the complement set. Without these explicit constraints, subtours will appear in the solution space which seemingly satisfy all the other constraints and may have lower cost than the real solution. For example, in an 8-node, 2-vehicle problem there could be the set of routes $0 \rightarrow 1 \rightarrow 2 \rightarrow 0$, $0 \rightarrow 3 \rightarrow 4 \rightarrow 0$, and a subtour $5 \rightarrow 6 \rightarrow 7 \rightarrow 5$. This set satisfies all constraints except subtour elimination. This issue has been noted in previous works such as \cite{azad2022solving}, which saw infeasible solutions appear in simulated QAOA output as their formulation did not include penalties for subtours.

Alternatively, route based formulations of the VRP assign a binary variable, $x_r$, to each possible route on the network, and the problem can be written as

$$\begin{aligned} \min _{\mathbf{x}} & \sum_{r \in \mathcal{R}} c_r x_r \\ \text { s.t. } & \sum_{r \in \mathcal{R}} \delta_{i, r} x_r=1 \quad \forall i \in \mathcal{N} \\ & x_r \in\{0,1\} \quad \forall r \in \mathcal{R}\end{aligned}$$

where $c_r$ is the cost of a route, $\delta_{i,r}$ is 1 if route $r$ visits node $i$, and $\mathcal{R}$ is the set of routes. While this formulation has fewer constraints, the number of routes in a fully connected network is $|\mathcal{N}|!$, which have to be generated explicitly in a preprocessing step before any optimization algorithm is run. Since each route corresponds to a qubit in this scheme, it would require a factorial number of qubits to implement making it infeasible except for very small problems.  

Another encoding method uses the time step expansion by including time steps into the decision variable $x^k_{v.t}$, which is $1$ if the $k$th vehicle visits the node $v$ at time step $t$. This is also known as a sequence-based formulation~\cite{harwood2021formulating}. The optimization problem is then 
$$\min _{\mathbf{x}} \sum_{v, v^{\prime}, t, k} w_{v v^{\prime}} \cdot x_{v, t}^k \cdot x_{v^{\prime}, t+1}^k$$
$$\begin{aligned} \text{s.t. } \   & \sum_{v=1}^n\left(1-\sum_{k=1}^K \sum_{t=1}^T x_{v, t}^k\right)^2 = 0 \\  & \sum_{k=1}^K \sum_{t=1}^T\left(1-\sum_{v=0}^n x_{v, t}^k\right)^2 = 0 \\ & \sum_{t=1}^T \sum_{v=1}^n x_{v, t}^k \cdot d_v \leq C \quad \forall k \in K\end{aligned}$$

where $d_v$ is the demand of a node and $C$ is the capacity of a vehicle. The assumption of demands always being met simplifies the last constraint, but the qubit requirements for this formulation are $O(n \cdot T \cdot K)$. This method was used in \cite{palackal2023quantum} with a clustering preprocessing phase to simplify the problem. 

A similar formulation is adopted in \cite{fitzek2024applying} by introducing a routing variable that tracks the order in which each node is visited, eliminating the need to consider subtour constraints. In the Results section we compare the resource requirements for these formulations. For now, we continue with the link (edge)-based formulation discussed first.

\section{Experiment Design and Implementation}
\begin{figure*}[!ht]
    \centering
    \includegraphics[width=1\linewidth]{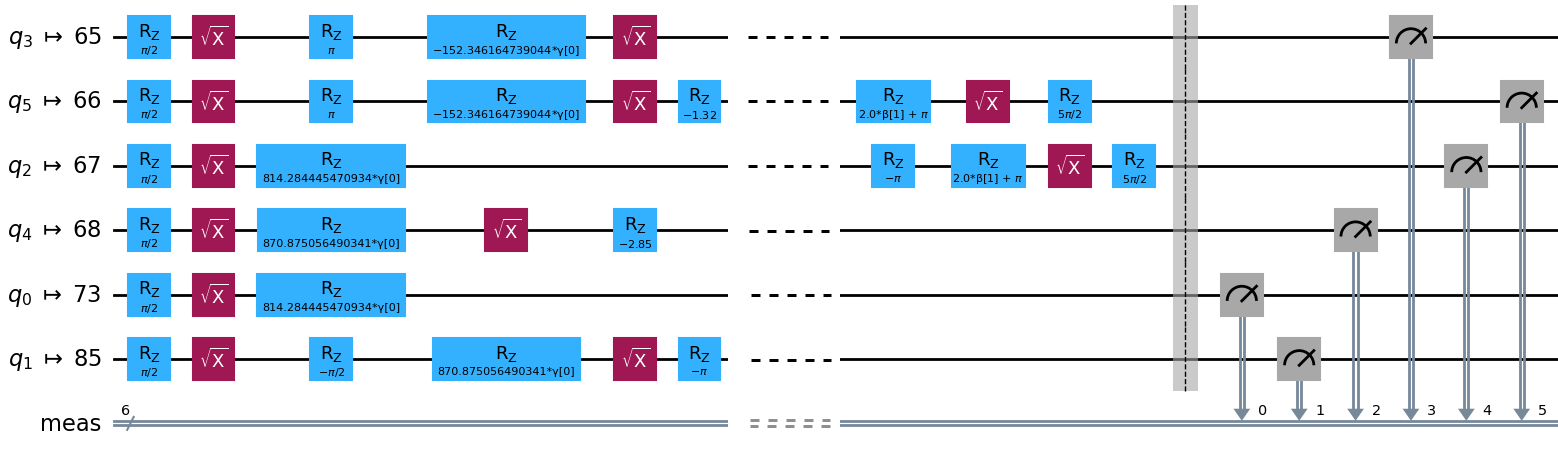}
    \caption{QAOA circuit with the left part showing the mapping to physical qubits on the quantum computer; the initialization $R_Z(\pi/2)$ refers to the Hadamard gate for superposition. On the right is the measurement of each qubit. The middle part has been truncated for space. }
    \label{fig:3qaoacircuit}
\end{figure*}

The QAOA was formulated from a QUBO representing a VRP instance based initially on a qiskit community example~\cite{vrpQiskit}, and implemented on the 127 qubit IBM-Rensselaer quantum computer. An excerpt of the transpiled parametrized circuit is depicted in Figure \ref{fig:3qaoacircuit}. This parameters were optimized classically, and the final circuit was run to sample the solution.

\subsection{QAOA Formulation}
A randomly instantiated VRP with 3 nodes and 2 vehicles was formulated using the link based formulation in Equation~\ref{eq:vrp} as a combinatorial optimization problem with a linear cost function and linear constraints. Qiskit optimization libraries were used to create the instance of the problem and convert to QUBO and the Ising Hamiltonian. While this package is no longer officially supported by IBM, the functionality may be adapted to work with the latest version of Qiskit. The coefficients of the following instance are the squared euclidean distance between the nodes indicated by the subscripts.

Minimize:
\begin{equation} \label{3nodeEQ}
\begin{split}
    \sum_{i \sim j} w_{i,j} x_{i,j} = \  &61.323  x_{0,1} + 4.732  x_{0,2} + 61.323  x_{1,0} \\
     + &42.895  x_{1,2} + 4.732  x_{2,0} + 42.895  x_{2,1}
\end{split}
\end{equation}
Subject to:
\begin{align*}
     x_{1,0} + x_{1,2} &= 1 \quad & (C_0) \\
     x_{2,0} + x_{2,1} &= 1 \quad & (C_1) \\
     x_{0,1} + x_{2,1} &= 1 \quad & (C_2) \\
     x_{0,2} + x_{1,2} &= 1 \quad & (C_3) \\
     x_{1,0} + x_{2,0} &= 2 \quad & (C_4) \\
     x_{0,1} + x_{0,2} &= 2 \quad & (C_5) \\
     x_{1,2} + x_{2,1} &\leq 1 \quad & (C_6) \\
     x_{i,j} \in \{0,1&\} \ \forall i,j \quad & (C_7)
\end{align*}

Constraints $C_0$ and $C_1$ enforce that a vehicle may leave one node only once, constraints $C_2$ and $C_3$ say that a vehicle may enter a node only once, $C_4$ states that there are two vehicles entering the depot, and $C_5$ requires that two vehicles leave the depot. Finally, $C_6$ eliminates sub-tours by saying that the link between nodes 1 and 2 may be utilized once or not at all, but not more than that. This removes the possibility of a vehicle going from node 1 to 2 then back 2 to 1. The statement $C_7$ represents that each variable $x_{i,j}$ is a binary number. The notation $i \sim j$ refers to valid connections in a graph. In this case we assume all links are valid. 

The formulation was converted to a quadratic unconstrained binary optimization (QUBO) problem by creating an augmented objective function similar to the Lagrange multiplier approach~\cite{glover2018tutorial}. This method adds to the original cost function quadratic penalties corresponding to each constraint, resulting in an unconstrained problem which has the same global optimum as the original problem. For example, $C_0$ can be converted to a penalty term $P(1-x_{1,0}-x_{1,2}+2x_{1,0}x_{1,2})$, where $P$ is a sufficiently large positive scalar, and $C_6$ may be added as $Px_{1,2}x_{2,1}$. Note that these transformations are unique to binary variables. Inequality constraints involving more than two binary variables typically require the introduction of auxiliary slack variables to be represented exactly as quadratic penalty terms.

The specific value of $P$ should be high enough that a constraint violation results in a significant penalty as compared to the original cost function, but not too high that it hinders the optimization process. For fine-tuning purposes each constraint may even have a different penalty value to reflect its importance. As a rule of thumb, $P>\sum_{i\sim j} |w_{i,j}|$ is considered sufficient \cite{harwood2021formulating}, but higher values are deemed more practical to aid the optimization algorithm. In this study a value of $2\sum_{i\sim j} |w_{i,j}|$ was used for all penalties. 

The augmented cost function for this problem can then be written as the following QUBO. 
\begin{equation}\begin{split}
&\min \mathbf{x}^T\mathbf{Qx} +b  = \\
&875.607 \, x_{0,1} x_{0,2} + 875.607 \, x_{0,2} x_{1,2} + 875.607 \, x_{1,0}x_{1,2} \\ 
&+ 875.607 \, x_{1,0} x_{2,0} + 875.607 \, x_{0,1} x_{2,1} + 218.901 \, x_{1,2} x_{2,1} \\
&+ 875.607 \, x_{2,0} x_{2,1} - 1689.892 \, x_{0,1} - 1746.482 \, x_{0,2} \\
&- 1689.892 \, x_{1,0} - 832.712 \, x_{1,2} - 1746.482 \, x_{2,0} \\ 
&- 832.712 \, x_{2,1} + 5253.645 \end{split}\end{equation}

The constant term in this formulation is of no consequence and can be freely dropped, but is maintained here for consistency in the cost function value. Due to the binary nature of the variables, quadratic terms of the form $x_{i,j}^2$ turn into linear terms since $x^2 = x$ for $x \in \{0,1\}$. 

The binary variables are directly mapped to qubits, and each linear and quadratic term is then converted to an $N$-qubit operator (here $N$ = 6) in Pauli basis representation. Linear terms result in one Z operator on the corresponding qubit, while quadratic terms form commuting Z terms which correspond to controlled Z gates on the quantum computer. The resulting Hamiltonian from equation \ref{eq:Ising}, for our problem, then becomes
\begin{equation} \label{hamiltonian} \begin{split}
H_C &= 407.14 \,Z_5 + 435.43 \,Z_4 + 407.14 \,Z_3 - 76.17 \,Z_2 \\
&+ 435.43\, Z_1 - 76.17 \,Z_0 + 218.9\, Z_4 Z_5 + 218.9 \, Z_0 Z_5 \\
&+ 218.9 \,Z_2 Z_4 + 218.9 \,Z_2 Z_3 + 218.9 \,Z_1 Z_3 \\
&+ 54.72 \,Z_0 Z_2 + 218.9 \,Z_0 Z_1
\end{split}
\end{equation}
where $[Z_0,Z_1,Z_2,Z_3,Z_4,Z_5]$  respectively correspond to $[x_{0,1}, x_{0,2}, x_{1,0}, x_{1,2}, x_{2,0}, x_{2,1}]$.

Each term in the Hamiltonian is effectively an operation on all qubits, represented as a sparse Pauli operation in Qiskit which are stored as a list of Pauli basis components. For example $Z_2$ would correspond to \verb|IIZIII|, and $Z_1Z_3$ corresponds to \verb|IZIZII|, where \verb|I| is the identity operation. 

The mixer Hamiltonian, $H_M$, is chosen to be the commonly used $H_M  = X^{\otimes n} $, where $X$ is  the Pauli-$X$ gate.

\subsection{Quantum Circuit Synthesis}
With the cost Hamiltonian prepared, the QAOA quantum circuit can be created. In this example, we choose a depth of 2, which means it will have the parameters $\gamma_0, \beta_0, \gamma_1, \beta_1$. The initial value for $\gamma$ is chosen to be $\pi$ and for $\beta$ it is taken to be $\frac{\pi}{2}$. The circuit is generated using Qiskit library provided QAOA ansatz and transpiled to the IBM-Rensselaer quantum computer. The beginning and end of the circuit is shown in Figure \ref{fig:3qaoacircuit}. It is composed of 213 $R_Z$ gates and 104 $R_X$, 43 echoed cross-resonance gates (CNOT), and 11 Pauli-X gates.

A cost function estimator was constructed to run the circuit with the given $\gamma$s and $\beta$s, to output the cost. Around this estimator was wrapped a classical optimizer known as Constrained Optimization by Quadratic Approximation (COBYQA) to iteratively find new values of $\gamma$ and $\beta$ which ideally bring the expected value of the cost, $\langle\psi|H_C|\psi\rangle$, down. 

Error suppression was applied to the circuit in the form of dynamic decoupling, which maintains coherence on idle qubits that can pick up unwanted interference from other qubits. This is a hardware feature that works by sending precisely timed pulses that amount to an identity operation on the qubit \cite{ezzell2023dynamical}. Pauli twirling is a technique used to turn arbitrary noise channels into specific signatures, and was also used in the experiments. The method essentially introduces Pauli rotations that eventually cancel out, increasing the fidelity of the overall circuit \cite{wallman2016noise}.  

\begin{figure}[!htb]
    \centering
    \includegraphics[width=0.75\linewidth]{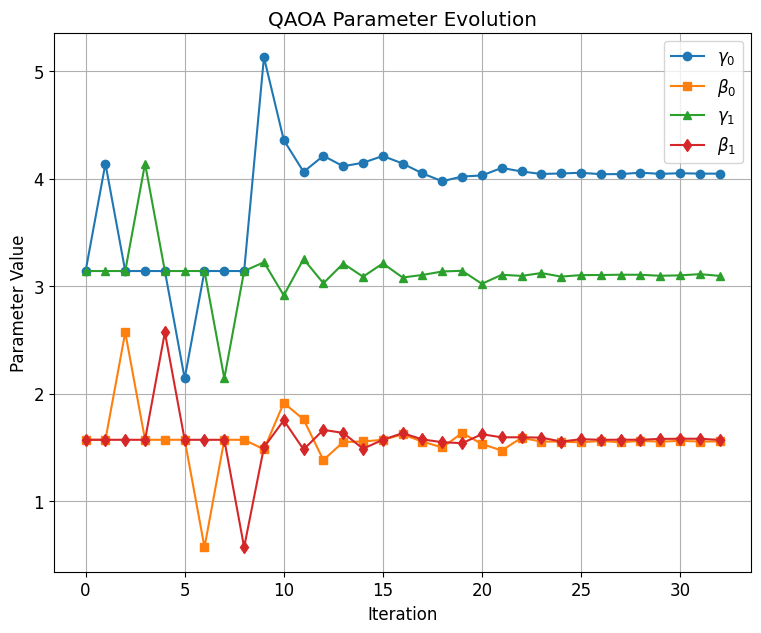}
    \caption{Convergence of QAOA parameters over optimization iterations}
    \label{fig:qaoaparams}
\end{figure}

The transpiled circuit was then used to run the hybrid QAOA on the IBM-Rennselaer Eagle r3 quantum computer, and the optimization process took around 14 minutes. After these values converged as depicted in Figure \ref{fig:qaoaparams}, the final parameters were assigned and the circuit executed again for 10,000 shots, which take about 15 seconds of wall clock time.

\subsection{Results and Interpretation}
\begin{figure*}[!htb]
    \centering
    \includegraphics[width=1\linewidth]{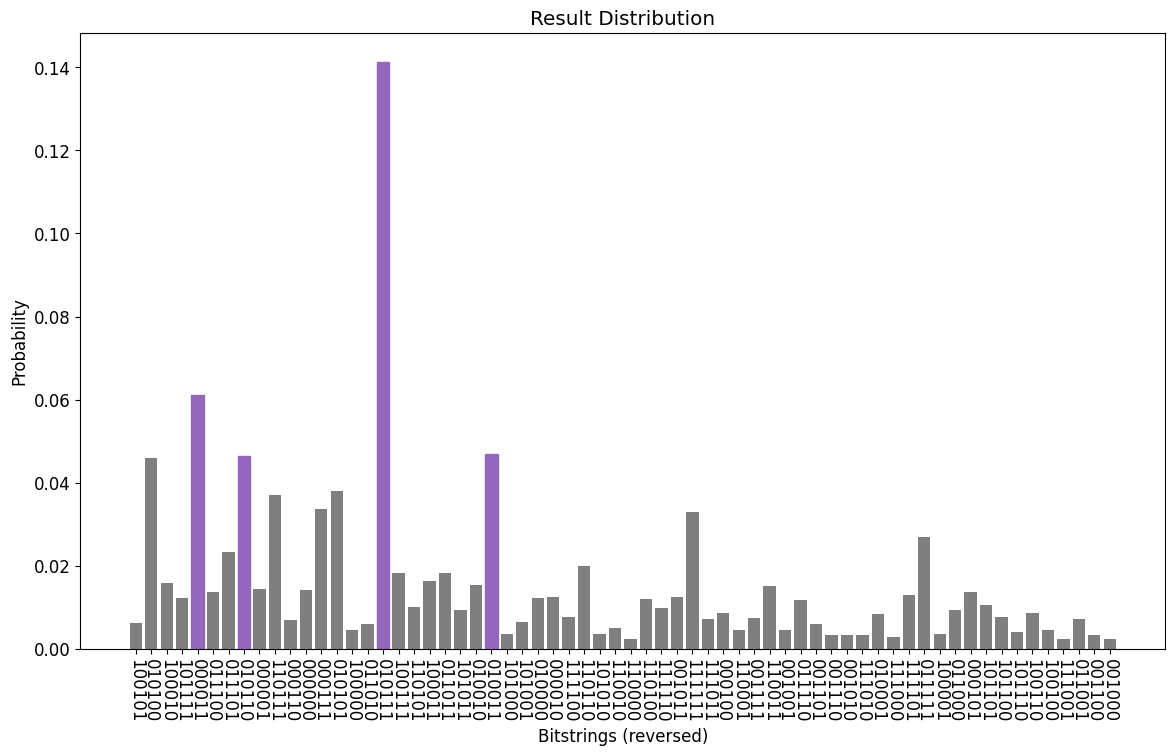}
    \caption{Sampled distribution from optimized circuit}
    \label{fig:result3}
\end{figure*}

The results obtained from running the hybrid QAOA algorithm for the VRP provide insights into the effectiveness and scalability of quantum computing for combinatorial optimization tasks. The final sampling using the trained parameters resulted in a distribution of results, shown in Figure~\ref{fig:result3}. It can be seen that the most likely bit-string is 111010, which corresponds to $x_{0,1}, x_{0,2}, x_{1,0}, x_{2,0}$ having the value 1, and the remaining variables are 0. It has been verified with a classical solver that this is the optimum solution. 

\begin{figure}[!htb]
    \centering
    \includegraphics[width=0.65\linewidth]{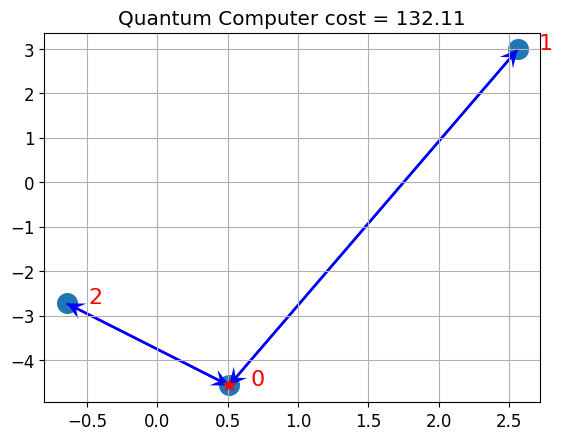}
    \caption{Assigned routes for the 3 node, 2 vehicle problem. Each arrow head indicates a directed edge used in the solution, and the red numbers are node labels.}
    \label{fig:routes3}
\end{figure}

This outcome suggests that the algorithm was able to identify the optimal solution with a high probability. However, the probability distribution also shows that suboptimal solutions were sampled with non-negligible frequencies, a behavior that could be attributed to noise in the quantum hardware, low depth $p$, and limitations of the standard QAOA. For finite $p$ QAOA is not expected to return the optimal solution with probability 1. 

The routing solution corresponding to the most probable bit-string is shown in Figure \ref{fig:routes3}. The three nodes are labeled in red, with 0 being the depot node. The selected routes (depicted by arrows) from QAOA successfully satisfy the constraints of the 3-node, 2 vehicle problem, demonstrating that QAOA was able to generate a feasible routing configuration.

\begin{figure}[!htpb]
    \centering
    \begin{subfigure}[b]{1\textwidth}
        \centering
        \includegraphics[width=0.69\linewidth]{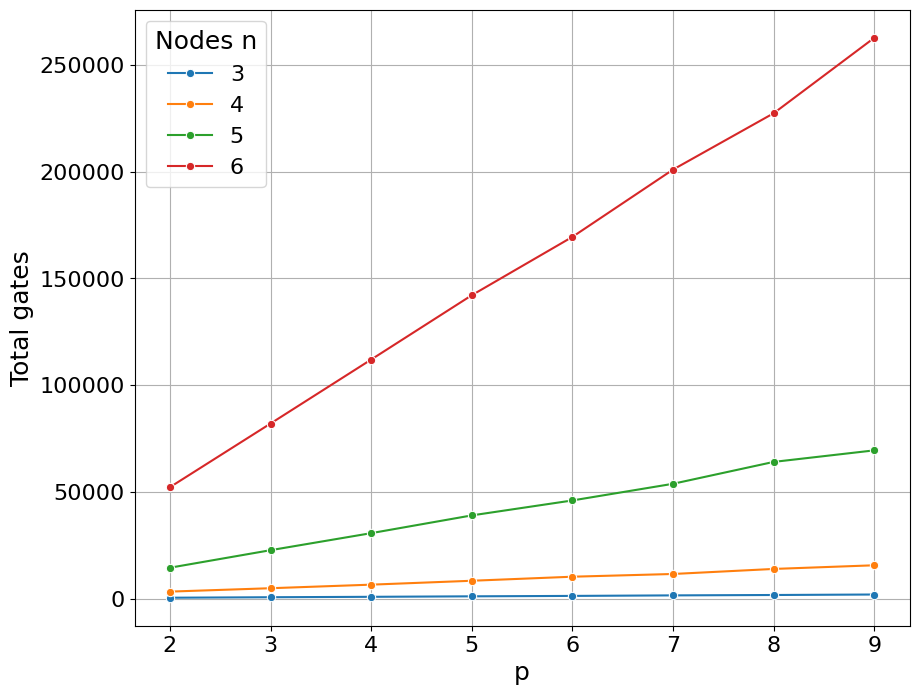}
        \caption{Problem scaling with $p$, for different values of $n$ \vspace{0.25cm} }
            \label{fig:gatesvsP}
    \end{subfigure}

    \begin{subfigure}[b]{1\textwidth}
        \centering
        \includegraphics[width=0.67\linewidth]{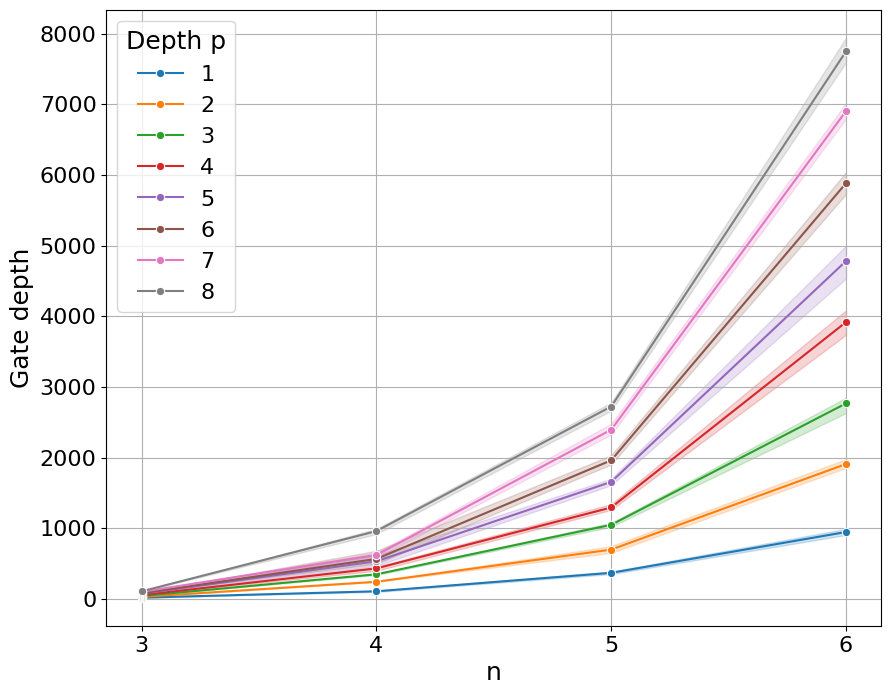}
        \caption{Circuit depth scaling with $n$, for different values of $p$. The shading indicates variance for different instances of the problem.}
        \label{fig:gatesvsN}
    \end{subfigure}
         
    \caption{Total circuit gates and circuit depth for different versions of the problem}
\end{figure}

Our findings also highlight the rapid growth in quantum gate counts as the problem scales. Larger instances that require more depth quickly become too large to run practically. For example a 4 node problem with depth of 2 generates a 14 qubit circuit consisting of 3207 $R_Z$, 1642 $R_X$, 928 echoed cross-resonance, and 214 Pauli-X gates. This problem took over 2 hours to run on the quantum computer and generated an infeasible solution. In fact, the top 10 most sampled states were all infeasible. Similarly, a 5 node problem with depth 2 transpiled into a 30 qubit circuit of about 30,000 basic gates. The number of circuit gates increases linearly with $p$ as plotted in Figure \ref{fig:gatesvsP}, while the total gates increases exponentially against the number of nodes in the problem. However, a more representative measure of circuit depth from a noise accumulation perspective is the maximum number of two qubit gates in a path from the start of the circuit to the end, shown in Figure \ref{fig:gatesvsN}. Using leave-one-out cross-validation on four problem sizes, a quadratic fit exhibits lower prediction error than an exponential one, indicating that the observed growth is consistent with quadratic scaling over the tested range.

\subsection{Circuit complexity}

QAOA scales quadratically with the number of qubits, and linearly with depth. If there are $n$ qubits, each cost unitary Hamiltonian contains $O(n^2)$ terms as can be seen from Equation \ref{eq:Hising}. Each $Z_i$ can be implemented with a single $R_z$ gate, while the $Z_iZ_j$ terms require two CNOTs and one $R_z$ gate. Meanwhile, the mixing Hamiltonian is implemented with $n$ $R_x$ gates. Therefore, each of the $p$ layers contains $O(n^2)$ gates, resulting in the overall circuit. 

\begin{figure}
    \centering
    \includegraphics[width=0.7\linewidth]{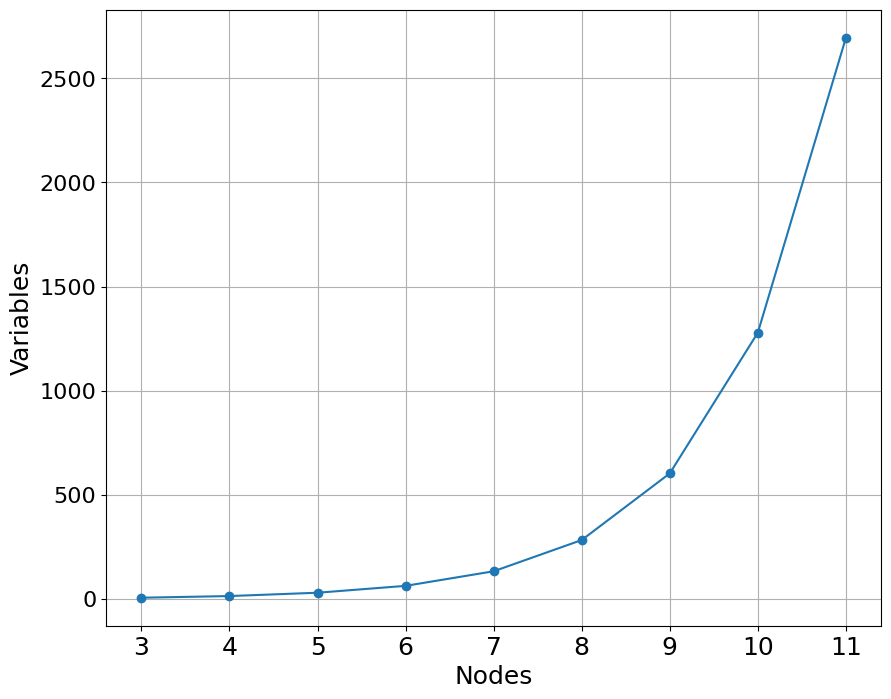}
    \caption{Number of variables in the QUBO for a few instances of the VRP}
    \label{fig:qubovars}
\end{figure}

Using the edge-based formulation, the main complexity in the case of the VRP arises from the subtour elimination constraints, which increase exponentially with the number of nodes, since there is one constraint for every subset of nodes and $O(2^n)$ subsets. Each constraint with more than 3 variables can be broken into multiple smaller constraints by introducing new binary variables, which is the approach used for generating the QUBO. As a result, there ends up being an exponentially increasing number of terms in the QUBO and, therefore, the Ising Hamiltonian. To illustrate, the VRP was instantiated for a number of nodes, and the number of variables in the QUBO were counted to plot the graph in  Figure~\ref{fig:qubovars}.

\subsubsection{Choice of Routing Formulation}
Among the three routing formulations examined earlier, we selected the edge-based formulation with subtour elimination for quantum implementation. Although time-expanded and order-based encodings remove subtours structurally, their substantially larger QUBO sizes and resulting two-qubit gate depths render them impractical on current quantum hardware. 

\begin{figure}[!htpb]
    \centering
    \begin{subfigure}[b]{1\textwidth}
        \centering
        \includegraphics[width=0.69\linewidth]{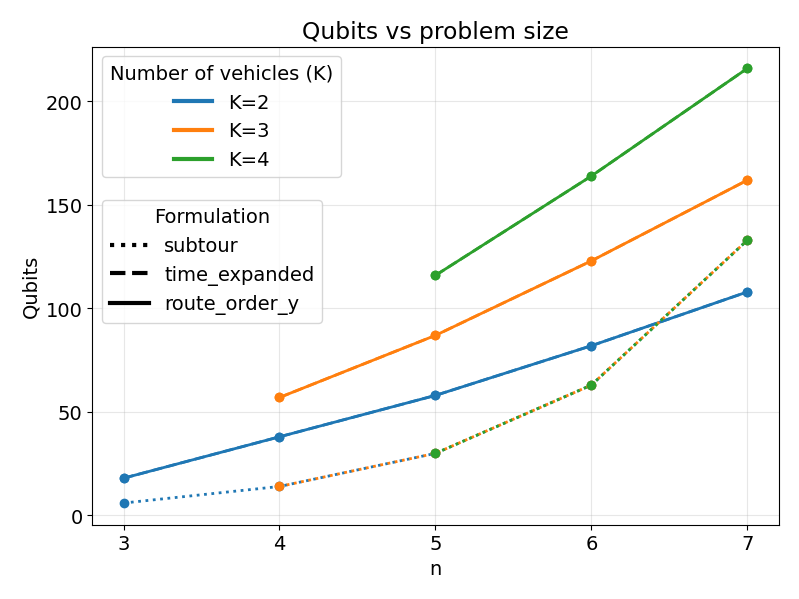}
        \caption{Number of qubits vs.  problem size for different number of vehicles}
            \label{fig:formulationN_q_n}
    \end{subfigure}

    \begin{subfigure}[b]{1\textwidth}
        \centering
        \includegraphics[width=0.67\linewidth]{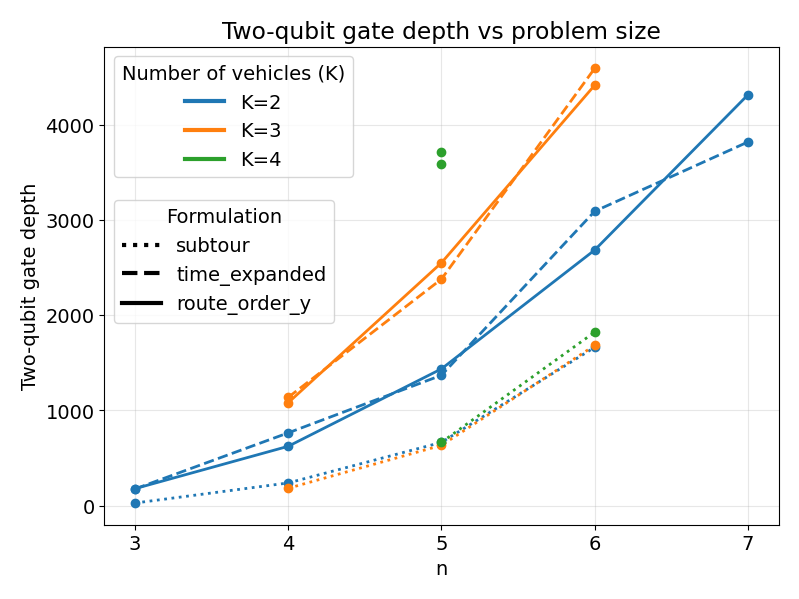}
        \caption{Two-qubit gate depth vs. problem size for different number of vehicles}
        \label{fig:formulation_2qdepth_n}
    \end{subfigure}
         
    \caption{Comparison of qubits and two-qubit gate depth for edge-based vs. time-expanded \cite{palackal2023quantum} and route-order~\cite{fitzek2024applying} formulations. $p=2$ for all cases.}
\end{figure}

The first issue is that of the number of qubits. The IBM-Rensselaer quantum computer has 127 qubits, which puts a limit on the size of any problem that can be executed on it. Figure \ref{fig:formulationN_q_n} shows the number of qubits needed to run the formulations discussed in Section 3.3, highlighting the multiplicative effect of introducing routing variables for each vehicle. The time-expanded and route-order formulations require the same number of binary variables, therefore their graphs overlap. Note that we do not consider vehicle-flow  formulations as they are more relevant when demands and capacities are considered \cite{irnich2014chapter}. For example, the two-index flow formulation from \cite{munari2016generalized} produces a QUBO requiring more variables than any of the other methods. 

Figure \ref{fig:formulation_2qdepth_n} shows the two-qubit gate depth for the same formulations after transpilation to our target hardware. For four vehicles there are two isolated points corresponding to the time-expanded and route-ordered formulations in the 5-node case, since there cannot be four vehicles in a smaller problem, and for larger problems the number of qubits required is greater than the quantum computer has available, so transpilation fails. The edge-based formulation requires a lower two-qubit gate depth across all cases.

Consequently, by both practical measures the edge-based formulation provides the most feasible way to execute QAOA for problems in this regime on available hardware.

\subsection{Penalty Parameter Tuning and Normalization Effects}

We investigated the impact of penalty scaling on solution quality. Choosing either substantially higher or lower penalty values did not improve performance. Instead, setting the penalty coefficient to twice the sum of edge weights was found to consistently yield the most reliable results.
In addition, we examined the effect of normalizing the coefficients of the Pauli terms so that all magnitudes were bounded by one. This normalization improved performance, particularly for larger problem instances with 4 and 5 nodes, where QAOA otherwise struggled to recover feasible solutions. 

With normalization applied, QAOA for the 4-node problem was able to recover the correct solution among the most frequently sampled bitstrings. Without normalization, however, the algorithm failed to consistently identify the optimal route in the high probability region of the output distribution. This observation was consistent across multiple randomly generated 4 and 5 node instances.

\begin{figure}
    \centering
    \includegraphics[width=0.75\linewidth]{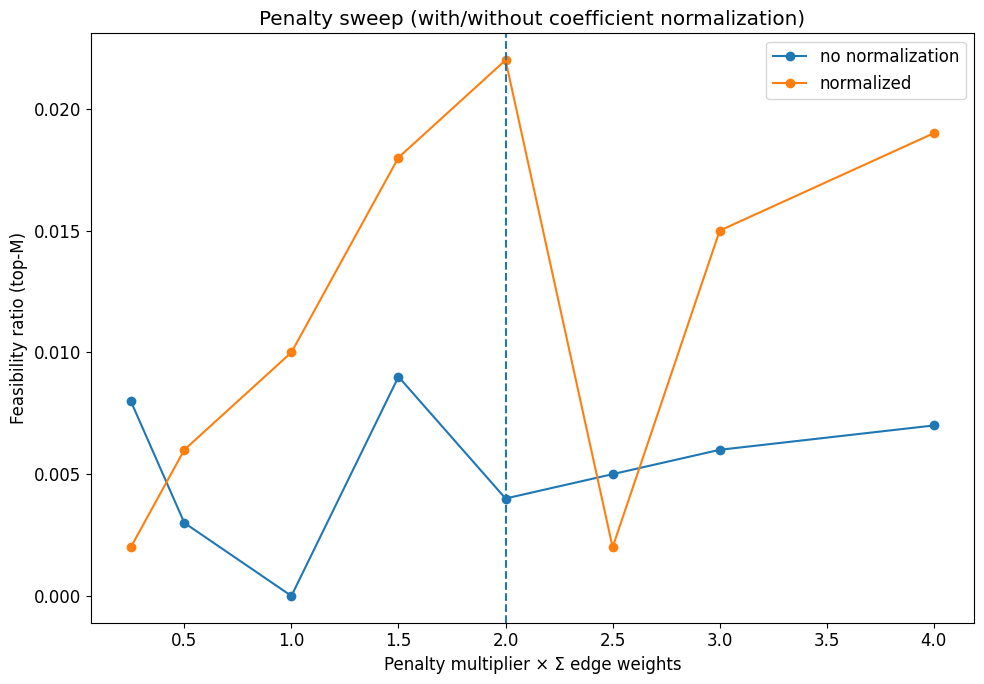}
    \caption{Effects of normalization and penalty multiplier on the 4 node VRP}
    \label{fig:penalty}
\end{figure}

Figure \ref{fig:penalty} shows the feasibility ratio of the top 1000 most frequently sampled bitstrings as a function of the penalty multiplier, with and without coefficient normalization. Without normalization, feasibility remains low across all penalty values, showing only small fluctuations. In contrast, normalization substantially improves performance, particularly around multipliers of 1.5--2.0, where the feasibility ratio peaks at over 2\%. Beyond this range, normalized performance remains consistently higher than the non-normalized case for almost all multipliers tested.

These results indicate two key trends: \textbf{(i)} scaling the penalty to approximately twice the sum of edge weights provides a robust choice across settings, and \textbf{(ii)} normalization of the Hamiltonian coefficients is critical for improving solution quality on larger problem instances.

Beyond demonstrating a single successful instance, the results provide actionable guidance on penalty selection, Hamiltonian normalization, and formulation choice that directly inform how constrained optimization problems should be encoded for near-term quantum processors.


\subsection{Limitations and Future Considerations}
While the QAOA has demonstrated potential for solving combinatorial optimization problems, as the problem size increases, it encounters several challenges. The observed exponential growth in circuit complexity caused by the problem mapping leads to significant decoherence and noise accumulation, reducing solution accuracy. Without constraints, the number of qubits would increase with the square of the number of nodes in the VRP because the problem requires representation of pairwise interactions between nodes. However, the inclusion of subtour elimination constraints adds an exponentially increasing number of slack variables and Pauli terms to the problem Hamiltonian.
While methods such as unbalanced penalization \cite{montanez2024unbalanced} can reduce the number of variables, they introduce additional interaction terms into the Hamiltonian, which may not be a favorable tradeoff given the overhead of swap and other two-qubit gates required to implement those interactions on physical quantum hardware. Moreover, the penalty coefficients would have to be optimized classically for every problem beforehand. Future quantum computers with more qubits and lower two-qubit error rates could take advantage of route-order formulations as they provide sub-exponential scaling, but this becomes favorable only for larger problems. 

Beyond current hardware limitations for all quantum computers in the NISQ era, the classical optimization component of QAOA, responsible for tuning the variational parameters $\beta$ and $\gamma$, becomes increasingly difficult as the VRP instance size grows, requiring more efficient initialization and hybrid optimization techniques. This parameter optimization requires navigating an increasingly complex cost function landscape, where gradient-based methods struggle with vanishing gradients and non-convexity. As $p \rightarrow \infty$, QAOA should theoretically find the solution with high probability, however, even with short depth QAOA has been found to offer significant advantages for various combinatorial optimization problems \cite{farhi2015bounded}. Recent studies have indicated that the critical circuit depth required to achieve a success probability of at least 70\% is highly dependent on the problem size and constraint density \cite{akshay2022circuit}. This scaling presents a fundamental challenge, as larger circuits require longer coherence times and more precise gate operations.
To compound things further, the optimization landscape of QAOA becomes increasingly complex as the number of qubits grows, leading to an exponential increase in the number of function evaluations required for parameter tuning

To address these limitations, several alternative approaches have been explored. One promising avenue is the use of Quantum Random Access Codes to relax combinatorial optimization problems, as demonstrated with Max-Cut, where this encoding achieved an approximation ratio of 0.905 for 40-node graphs  \cite{fuller2024approximate}. 
While encoding schemes have been used on the VRP to minimize the number of qubits needed,  they introduce additional trade-offs. The encoded states are not able to fully capture the classical correlation between variables leading to much larger variance during the optimization process \cite{leonidas2023qubit}. 

A relevant critical insight from recent research is that QAOA parameters $\gamma$ and $\beta$ lie in concentrated regions, and can be precomputed using smaller problem instances. This enables efficient scaling by requiring only polynomial-time training to obtain the optimal values for larger systems \cite{akshay2021parameter}. Another study has found that the parameters follow a linear relationship with respect to circuit depth $p$, so for a given problem instance, an initial coarse optimization of only four parameters may be sufficient before refining with a full $2p$ parameter optimization~\cite{sakai2024linearly}. Moreover, these optimized parameters have been found to transfer between problem instances, further enhancing computational efficiency.
In a related strategy, a smaller circuit can be used to optimize a subset of the parameters, which can then be transferred to the larger circuit requiring optimization on only a small subset of new parameters \cite{venturelli2024investigating}. This method allows optimization to focus only on a small subset of new parameters, significantly reducing the overall computational burden while maintaining solution quality.

Deep quantum circuits face inherent limitations due to cumulative gate errors and decoherence, which degrade computational accuracy as circuit depth increases. This necessitates robust error mitigation strategies to enhance the reliability of quantum computations. One promising approach is the use of dynamic circuits, which exploit intermediate measurements on ancilla qubits in conjunction with conditional feed-forward operations. By adaptively modifying the quantum state based on measurement outcomes, dynamic circuits can effectively reduce error accumulation and improve overall circuit fidelity. Recent experimental results on systems with up to 101 qubits have demonstrated improved fidelity in CNOT gate teleportation using such dynamic circuit techniques~\cite{baumer2024efficient}. 

While error mitigation and correction techniques may be applied to prevent decoherece, they often present a significant overhead, causing longer execution times. Operator backpropagation has been proposed recently to decompose the circuit into two subcircuits, one of which can be simulated efficiently on a classical computer. This technique effectively redistributes computational complexity, reducing quantum circuit depth at the cost of increased classical post-processing overhead. This trade-off is favorable when the circuit contains a limited number of non-Clifford gates where classical simulation remains efficient \cite{fuller2025improved}. 

In addition to circuit-level optimizations, there remains significant room for exploration in the design of mixer Hamiltonians in QAOA. Specifically, it has been proposed to encode the constraints of the problem directly into the mixer to restrict the quantum state evolution to the feasible solution space. This constraint-aware mixer design has been applied to the capacitated VRP resulting in a polynomial number of gates, but the circuit requires multi-controlled Toffoli gates \cite{xie2024feasibility}. This gate is difficult to implement due to their high error rates and resource-intensive nature. The authors note that the bit flip noise in the quantum device may not preserve the feasible states, leading to degraded performance. Addressing these challenges requires further investigation into noise-resilient quantum gate implementations and alternative formulations of constraint-encoded mixers

\section{Conclusion}

Quantum computing is an emerging and exciting avenue of research that is now becoming available to tackle problems from a different perspective. In this work, we explored the application of QAOA to the VRP by formulating it as a QUBO problem by encoding constraints such as flow conservation and subtour elimination into the cost function. Our study examined the feasibility of executing QAOA-based VRP solutions on current quantum hardware, with a particular focus on the impact of problem size and formulation on circuit complexity and noise resilience. 

Starting with a three-node VRP, the QUBO was cast as an Ising Hamiltonian for QAOA with depth 2. The variational parameters were optimized using a classical optimizer by executing the circuit on the quantum computer multiple times, and the converged values were used to sample the outcome of the final circuit. The most probable observed state corresponded with the known correct solution to the problem indicating a successful result for QAOA. 

However, for larger instances of the problem, it was found that while QAOA can, in principle, provide approximate solutions to VRP, the rapid growth in circuit depth, specifically two-qubit gate depth, poses a significant practical challenge due to the limitations of noisy quantum processors. This suggests that the problem is dependent on hardware improvements and algorithmic optimizations, particularly in reducing the number of multi-qubit operations and mitigating noise effects. Our experiments show that coefficient normalization and carefully tuned penalty values are essential for improving feasibility in QAOA-based VRP. These heuristics helped recover feasible solutions more reliably for larger problems, highlighting the importance of parameter design in practical quantum optimization. Future work will focus on developing more efficient QUBO formulations and encoding schemes, employing other hybrid approaches such as VQE and operator backpropagation, transfer learning of the variational parameters, and dynamic circuits. 

The primary contribution of this work lies in its experimental, hardware-grounded analysis of QAOA for VRP. Rather than proposing a new algorithmic variant, we systematically expose the practical bottlenecks that arise when mapping constrained routing problems onto real quantum processors. These results clarify why many prior QAOA VRP approaches struggle to produce feasible solutions in practice, and they outline concrete methodological directions for advancing quantum optimization beyond simulation-only studies. As quantum hardware continues to advance, these improvements will be vital in determining whether QAOA and related quantum algorithms can provide a competitive advantage over classical optimization methods for real-world applications.


\begin{acks}
This work is funded through the IBM-RPI Future of Computing Research Collaboration.
\end{acks}

\bibliographystyle{ACM-Reference-Format}
\bibliography{sample-base}

@String{Computing = "Computing" }

@String{Computer = "{IEEE} Computer" }

@String{Springer = "Springer-Verlag" }

@article{glover2018tutorial,
  title={A tutorial on formulating and using QUBO models},
  author={Glover, Fred and Kochenberger, Gary and Du, Yu},
  journal={arXiv preprint arXiv:1811.11538},
  year={2018}
}

@article{harwood2021formulating,
  title={Formulating and solving routing problems on quantum computers},
  author={Harwood, Stuart and Gambella, Claudio and Trenev, Dimitar and Simonetto, Andrea and Bernal, David and Greenberg, Donny},
  journal={IEEE Transactions on Quantum Engineering},
  volume={2},
  pages={1--17},
  year={2021},
  publisher={IEEE}
}

@article{bremner2011classical,
  title={Classical simulation of commuting quantum computations implies collapse of the polynomial hierarchy},
  author={Bremner, Michael J and Jozsa, Richard and Shepherd, Dan J},
  journal={Proceedings of the Royal Society A: Mathematical, Physical and Engineering Sciences},
  volume={467},
  number={2126},
  pages={459--472},
  year={2011},
  publisher={The Royal Society Publishing}
}

@article{shor1999polynomial,
  title={Polynomial-time algorithms for prime factorization and discrete logarithms on a quantum computer},
  author={Shor, Peter W},
  journal={SIAM review},
  volume={41},
  number={2},
  pages={303--332},
  year={1999},
  publisher={SIAM}
}

@inproceedings{grover1996fast,
  title={A fast quantum mechanical algorithm for database search},
  author={Grover, Lov K},
  booktitle={Proceedings of the twenty-eighth annual ACM symposium on Theory of computing},
  pages={212--219},
  year={1996}
}

@article{tacchino2020quantum,
  title={Quantum computers as universal quantum simulators: state-of-the-art and perspectives},
  author={Tacchino, Francesco and Chiesa, Alessandro and Carretta, Stefano and Gerace, Dario},
  journal={Advanced Quantum Technologies},
  volume={3},
  number={3},
  pages={1900052},
  year={2020},
  publisher={Wiley Online Library}
}

@article{zhou2020quantum,
  title={Quantum approximate optimization algorithm: Performance, mechanism, and implementation on near-term devices},
  author={Zhou, Leo and Wang, Sheng-Tao and Choi, Soonwon and Pichler, Hannes and Lukin, Mikhail D},
  journal={Physical Review X},
  volume={10},
  number={2},
  pages={021067},
  year={2020},
  publisher={APS}
}

@article{blekos2024review,
  title={A review on quantum approximate optimization algorithm and its variants},
  author={Blekos, Kostas and Brand, Dean and Ceschini, Andrea and Chou, Chiao-Hui and Li, Rui-Hao and Pandya, Komal and Summer, Alessandro},
  journal={Physics Reports},
  volume={1068},
  pages={1--66},
  year={2024},
  publisher={Elsevier}
}

@article{boulebnane2024solving,
  title={Solving boolean satisfiability problems with the quantum approximate optimization algorithm},
  author={Boulebnane, Sami and Montanaro, Ashley},
  journal={PRX Quantum},
  volume={5},
  number={3},
  pages={030348},
  year={2024},
  publisher={APS}
}

@article{montanez2024towards,
  title={Towards a universal QAOA protocol: Evidence of quantum advantage in solving combinatorial optimization problems},
  author={Montanez-Barrera, JA and Michielsen, Kristel},
  journal={arXiv preprint arXiv:2405.09169},
  year={2024}
}

@article{fuller2024approximate,
  title={Approximate solutions of combinatorial problems via quantum relaxations},
  author={Fuller, Bryce and Hadfield, Charles and Glick, Jennifer R and Imamichi, Takashi and Itoko, Toshinari and Thompson, Richard J and Jiao, Yang and Kagele, Marna M and Blom-Schieber, Adriana W and Raymond, Rudy and others},
  journal={IEEE Transactions on Quantum Engineering},
  year={2024},
  publisher={IEEE}
}

@article{farhi2016quantum,
  title={Quantum supremacy through the quantum approximate optimization algorithm},
  author={Farhi, Edward and Harrow, Aram W},
  journal={arXiv preprint arXiv:1602.07674},
  year={2016}
}

@article{farhi2014quantum,
  title={A quantum approximate optimization algorithm},
  author={Farhi, Edward and Goldstone, Jeffrey and Gutmann, Sam},
  journal={arXiv preprint arXiv:1411.4028},
  year={2014}
}

@article{farhi2020quantum,
  title={The quantum approximate optimization algorithm needs to see the whole graph: Worst case examples},
  author={Farhi, Edward and Gamarnik, David and Gutmann, Sam},
  journal={arXiv preprint arXiv:2005.08747},
  year={2020}
}

@article{akshay2021parameter,
  title={Parameter concentrations in quantum approximate optimization},
  author={Akshay, Vishwanathan and Rabinovich, Daniil and Campos, Ernesto and Biamonte, Jacob},
  journal={Physical Review A},
  volume={104},
  number={1},
  pages={L010401},
  year={2021},
  publisher={APS}
}

@article{venturelli2024investigating,
  title={Investigating layer-selective transfer learning of QAOA parameters for Max-Cut problem},
  author={Venturelli, Francesco Aldo and Das, Sreetama and Caruso, Filippo},
  journal={arXiv preprint arXiv:2412.21071},
  year={2024}
}

@article{baumer2024efficient,
  title={Efficient long-range entanglement using dynamic circuits},
  author={B{\"a}umer, Elisa and Tripathi, Vinay and Wang, Derek S and Rall, Patrick and Chen, Edward H and Majumder, Swarnadeep and Seif, Alireza and Minev, Zlatko K},
  journal={PRX Quantum},
  volume={5},
  number={3},
  pages={030339},
  year={2024},
  publisher={APS}
}

@article{sakai2024linearly,
  title={Linearly simplified QAOA parameters and transferability},
  author={Sakai, Ryo and Matsuyama, Hiromichi and Tam, Wai-Hong and Yamashiro, Yu and Fujii, Keisuke},
  journal={arXiv preprint arXiv:2405.00655},
  year={2024}
}

@article{koch2007charge,
  title={Charge-insensitive qubit design derived from the Cooper pair box},
  author={Koch, Jens and Yu, Terri M and Gambetta, Jay and Houck, Andrew A and Schuster, David I and Majer, Johannes and Blais, Alexandre and Devoret, Michel H and Girvin, Steven M and Schoelkopf, Robert J},
  journal={Physical Review A—Atomic, Molecular, and Optical Physics},
  volume={76},
  number={4},
  pages={042319},
  year={2007},
  publisher={APS}
}

@article{farhi2000quantum,
  title={Quantum computation by adiabatic evolution},
  author={Farhi, Edward and Goldstone, Jeffrey and Gutmann, Sam and Sipser, Michael},
  journal={arXiv preprint quant-ph/0001106},
  year={2000}
}

@article{pellow2024effect,
  title={The effect of classical optimizers and Ansatz depth on QAOA performance in noisy devices},
  author={Pellow-Jarman, Aidan and McFarthing, Shane and Sinayskiy, Ilya and Park, Daniel K and Pillay, Anban and Petruccione, Francesco},
  journal={Scientific reports},
  volume={14},
  number={1},
  pages={16011},
  year={2024},
  publisher={Nature Publishing Group UK London}
}

@article{mcclean2016theory,
  title={The theory of variational hybrid quantum-classical algorithms},
  author={McClean, Jarrod R and Romero, Jonathan and Babbush, Ryan and Aspuru-Guzik, Al{\'a}n},
  journal={New Journal of Physics},
  volume={18},
  number={2},
  pages={023023},
  year={2016},
  publisher={IOP Publishing}
}

@article{feynman1982simulating,
  title={Simulating Physics with Computers},
  author={Feynman, Richard P},
  journal={International Journal of Theoretical Physics},
  volume={21},
  number={6/7},
  year={1982}
}

@article{deutsch1985quantum,
  title={Quantum theory, the Church--Turing principle and the universal quantum computer},
  author={Deutsch, David},
  journal={Proceedings of the Royal Society of London. A. Mathematical and Physical Sciences},
  volume={400},
  number={1818},
  pages={97--117},
  year={1985},
  publisher={The Royal Society London}
}

@article{kirkpatrick1983optimization,
  title={Optimization by simulated annealing},
  author={Kirkpatrick, Scott and Gelatt Jr, C Daniel and Vecchi, Mario P},
  journal={science},
  volume={220},
  number={4598},
  pages={671--680},
  year={1983},
  publisher={American association for the advancement of science}
}

@article{finnila1994quantum,
  title={Quantum annealing: A new method for minimizing multidimensional functions},
  author={Finnila, Aleta Berk and Gomez, Maria A and Sebenik, C and Stenson, Catherine and Doll, Jimmie D},
  journal={Chemical physics letters},
  volume={219},
  number={5-6},
  pages={343--348},
  year={1994},
  publisher={Elsevier}
}

@article{martovnak2004quantum,
  title={Quantum annealing of the traveling-salesman problem},
  author={Marto{\v{n}}{\'a}k, Roman and Santoro, Giuseppe E and Tosatti, Erio},
  journal={Physical Review E—Statistical, Nonlinear, and Soft Matter Physics},
  volume={70},
  number={5},
  pages={057701},
  year={2004},
  publisher={APS}
}

@article{boothby2020next,
  title={Next-generation topology of d-wave quantum processors},
  author={Boothby, Kelly and Bunyk, Paul and Raymond, Jack and Roy, Aidan},
  journal={arXiv preprint arXiv:2003.00133},
  year={2020}
}

@article{feld2019hybrid,
  title={A hybrid solution method for the capacitated vehicle routing problem using a quantum annealer},
  author={Feld, Sebastian and Roch, Christoph and Gabor, Thomas and Seidel, Christian and Neukart, Florian and Galter, Isabella and Mauerer, Wolfgang and Linnhoff-Popien, Claudia},
  journal={Frontiers in ICT},
  volume={6},
  pages={13},
  year={2019},
  publisher={Frontiers Media SA}
}

@article{yarkoni2022quantum,
  title={Quantum annealing for industry applications: Introduction and review},
  author={Yarkoni, Sheir and Raponi, Elena and B{\"a}ck, Thomas and Schmitt, Sebastian},
  journal={Reports on Progress in Physics},
  volume={85},
  number={10},
  pages={104001},
  year={2022},
  publisher={IOP Publishing}
}

@article{leonidas2023qubit,
  title={Qubit efficient quantum algorithms for the vehicle routing problem on quantum computers of the nisq era},
  author={Leonidas, Ioannis D and Dukakis, Alexander and Tan, Benjamin and Angelakis, Dimitris G},
  journal={arXiv preprint arXiv:2306.08507},
  year={2023}
}

@article{zhuang2024quantum,
  title={Quantum computing in intelligent transportation systems: A survey},
  author={Zhuang, Yifan and Azfar, Talha and Wang, Yinhai and Sun, Wei and Wang, Xiaokun and Guo, Qianwen and Ke, Ruimin},
  journal={CHAIN},
  volume={1},
  number={2},
  pages={138--149},
  year={2024},
  publisher={Youke Publishing}
}

@article{burkacky2020will,
  title={Will quantum computing drive the automotive future},
  author={Burkacky, Ondrej and Pautasso, L and Mohr, N},
  journal={Mckinsey \& Company},
  volume={1},
  pages={33--38},
  year={2020}
}

@article{ruan2020quantum,
  title={The quantum approximate algorithm for solving traveling salesman problem},
  author={Ruan, Yue and Marsh, Samuel and Xue, Xilin and Liu, Zhihao and Wang, Jingbo and others},
  journal={Computers, Materials and Continua},
  volume={63},
  number={3},
  pages={1237--1247},
  year={2020},
  publisher={Tech Science Press}
}

@article{azad2022solving,
  title={Solving vehicle routing problem using quantum approximate optimization algorithm},
  author={Azad, Utkarsh and Behera, Bikash K and Ahmed, Emad A and Panigrahi, Prasanta K and Farouk, Ahmed},
  journal={IEEE Transactions on Intelligent Transportation Systems},
  volume={24},
  number={7},
  pages={7564--7573},
  year={2022},
  publisher={IEEE}
}

@article{theurich2021branch,
  title={A branch-and-bound approach for a Vehicle Routing Problem with Customer Costs},
  author={Theurich, Franziska and Fischer, Andreas and Scheithauer, Guntram},
  journal={EURO Journal on Computational Optimization},
  volume={9},
  pages={100003},
  year={2021},
  publisher={Elsevier}
}

@article{baker2003genetic,
  title={A genetic algorithm for the vehicle routing problem},
  author={Baker, Barrie M and Ayechew, MA1951066},
  journal={Computers \& Operations Research},
  volume={30},
  number={5},
  pages={787--800},
  year={2003},
  publisher={Elsevier}
}

@article{sicilia2016optimization,
  title={An optimization algorithm for solving the rich vehicle routing problem based on Variable Neighborhood Search and Tabu Search metaheuristics},
  author={Sicilia, Juan Antonio and Quemada, Carlos and Royo, Beatriz and Escuin, David},
  journal={Journal of Computational and Applied Mathematics},
  volume={291},
  pages={468--477},
  year={2016},
  publisher={Elsevier}
}

@article{fitzek2024applying,
  title={Applying quantum approximate optimization to the heterogeneous vehicle routing problem},
  author={Fitzek, David and Ghandriz, Toheed and Laine, Leo and Granath, Mats and Kockum, Anton Frisk},
  journal={Scientific Reports},
  volume={14},
  number={1},
  pages={25415},
  year={2024},
  publisher={Nature Publishing Group UK London}
}

@article{fuller2025improved,
  title={Improved Quantum Computation using Operator Backpropagation},
  author={Fuller, Bryce and Tran, Minh C and Lykov, Danylo and Johnson, Caleb and Rossmannek, Max and Wei, Ken Xuan and He, Andre and Kim, Youngseok and Vu, DinhDuy and Sharma, Kunal and others},
  journal={arXiv preprint arXiv:2502.01897},
  year={2025}
}

@article{akshay2022circuit,
  title={Circuit depth scaling for quantum approximate optimization},
  author={Akshay, Vishwanathan and Philathong, H and Campos, E and Rabinovich, Daniil and Zacharov, Igor and Zhang, Xiao-Ming and Biamonte, Jacob D},
  journal={Physical Review A},
  volume={106},
  number={4},
  pages={042438},
  year={2022},
  publisher={APS}
}

@misc{farhi2015bounded,
      title={A Quantum Approximate Optimization Algorithm Applied to a Bounded Occurrence Constraint Problem}, 
      author={Edward Farhi and Jeffrey Goldstone and Sam Gutmann},
      year={2015},
      eprint={1412.6062},
      archivePrefix={arXiv},
      primaryClass={quant-ph},
      url={https://arxiv.org/abs/1412.6062}, 
}

@article{preskill2018quantum,
  title={Quantum computing in the NISQ era and beyond},
  author={Preskill, John},
  journal={Quantum},
  volume={2},
  pages={79},
  year={2018}
}

@article{clarke1964scheduling,
  title={Scheduling of vehicles from a central depot to a number of delivery points},
  author={Clarke, Geoff and Wright, John W},
  journal={Operations research},
  volume={12},
  number={4},
  pages={568--581},
  year={1964},
  publisher={Informs}
}

@article{xie2024feasibility,
  title={A feasibility-preserved quantum approximate solver for the capacitated vehicle routing problem},
  author={Xie, Ningyi and Lee, Xinwei and Cai, Dongsheng and Saito, Yoshiyuki and Asai, Nobuyoshi and Lau, Hoong Chuin},
  journal={Quantum Information Processing},
  volume={23},
  number={8},
  pages={291},
  year={2024},
  publisher={Springer}
}

@article{NASA_Science_2024, title="NASA, partners target megacities carbon emissions", url={https://science.nasa.gov/earth/climate-change/greenhouse-gases/nasa-partners-target-megacities-carbon-emissions/}, journal={NASA Science}, author={NASA}, year={2024}, month={Oct}}

@article{he2023brownian,
  title={Brownian bridge-based speed imputation technique for truck energy consumption and emissions estimation},
  author={He, Xiaozheng and Wei, Yu and Holguin-Veras, Jose},
  journal={Transportation Research Part D: Transport and Environment},
  volume={114},
  pages={103546},
  year={2023},
  publisher={Elsevier}
}

@article{junger1995traveling,
  title={The traveling salesman problem},
  author={J{\"u}nger, Michael and Reinelt, Gerhard and Rinaldi, Giovanni},
  journal={Handbooks in operations research and management science},
  volume={7},
  pages={225--330},
  year={1995},
  publisher={Elsevier}
}

@article{josephson1964coupled,
  title={Coupled superconductors},
  author={Josephson, BD},
  journal={Reviews of Modern Physics},
  volume={36},
  number={1},
  pages={216},
  year={1964},
  publisher={APS}
}

@article{ezzell2023dynamical,
  title={Dynamical decoupling for superconducting qubits: a performance survey},
  author={Ezzell, Nic and Pokharel, Bibek and Tewala, Lina and Quiroz, Gregory and Lidar, Daniel A},
  journal={Physical Review Applied},
  volume={20},
  number={6},
  pages={064027},
  year={2023},
  publisher={APS}
}

@article{wallman2016noise,
  title={Noise tailoring for scalable quantum computation via randomized compiling},
  author={Wallman, Joel J and Emerson, Joseph},
  journal={Physical Review A},
  volume={94},
  number={5},
  pages={052325},
  year={2016},
  publisher={APS}
}

@misc{vrpQiskit,
  author={Qiskit Community },
  title={Vehicle Routing},
  year={2024},
  url={https://github.com/qiskit-community/qiskit-optimization/blob/stable/0.6/docs/tutorials/07_examples_vehicle_routing.ipynb},
}

@article{kluber2023trotterization,
  title={Trotterization in quantum theory},
  author={Kluber, Grant},
  journal={arXiv preprint arXiv:2310.13296},
  year={2023}
}

@inproceedings{palackal2023quantum,
  title={Quantum-assisted solution paths for the capacitated vehicle routing problem},
  author={Palackal, Lilly and Poggel, Benedikt and Wulff, Matthias and Ehm, Hans and Lorenz, Jeanette Miriam and Mendl, Christian B},
  booktitle={2023 IEEE International Conference on Quantum Computing and Engineering (QCE)},
  volume={1},
  pages={648--658},
  year={2023},
  organization={IEEE}
}

@article{montanez2024unbalanced,
  title={Unbalanced penalization: A new approach to encode inequality constraints of combinatorial problems for quantum optimization algorithms},
  author={Monta{\~n}ez-Barrera, Jhon Alejandro and Willsch, Dennis and Maldonado-Romo, Alberto and Michielsen, Kristel},
  journal={Quantum Science and Technology},
  volume={9},
  number={2},
  pages={025022},
  year={2024},
  publisher={IOP Publishing}
}

@article{elsharkawy2025integration,
  title={Integration of quantum accelerators with high performance computing—a review of quantum programming tools},
  author={Elsharkawy, Amr and To, Xiao-Ting Michelle and Seitz, Philipp and Chen, Yanbin and Stade, Yannick and Geiger, Manuel and Huang, Qunsheng and Guo, Xiaorang and Ansari, Muhammad Arslan and Mendl, Christian B and others},
  journal={ACM Transactions on Quantum Computing},
  volume={6},
  number={3},
  pages={1--46},
  year={2025},
  publisher={ACM New York, NY}
}

@incollection{irnich2014chapter,
  title={Chapter 1: The family of vehicle routing problems},
  author={Irnich, Stefan and Toth, Paolo and Vigo, Daniele},
  booktitle={Vehicle Routing: Problems, Methods, and Applications, Second Edition},
  pages={1--33},
  year={2014},
  publisher={SIAM}
}

@article{munari2016generalized,
  title={A generalized formulation for vehicle routing problems},
  author={Munari, Pedro and Dollevoet, Twan and Spliet, Remy},
  journal={arXiv preprint arXiv:1606.01935},
  year={2016}
}










\end{document}